\documentclass[prb,twocolumn,superscriptaddress,showkeys,floatfix,altaffilsymbol]{revtex4-1}
\usepackage{graphicx}
\usepackage{amsmath}
\begin{document}

\title{Physical and magnetic properties of Ba(Fe$_{1-x}$Mn$_x$)$_2$As$_2$ single crystals}

\author{A. Thaler}
\affiliation{Ames Laboratory and Department of Physics and Astronomy, Iowa State University, Ames, IA 50011, USA}
\author{H. Hodovanets}
\affiliation{Ames Laboratory and Department of Physics and Astronomy, Iowa State University, Ames, IA 50011, USA}
\author{M. S. Torikachvili}
\affiliation{Ames Laboratory and Department of Physics and Astronomy, Iowa State University, Ames, IA 50011, USA}
\affiliation{Department of Physics, San Diego State University, San Diego, CA 92182}
\author{S. Ran}
\affiliation{Ames Laboratory and Department of Physics and Astronomy, Iowa State University, Ames, IA 50011, USA}
\author{A. Kracher}
\affiliation{Ames Laboratory and Department of Physics and Astronomy, Iowa State University, Ames, IA 50011, USA}
\author{W. Straszheim}
\affiliation{Ames Laboratory and Department of Physics and Astronomy, Iowa State University, Ames, IA 50011, USA}
\author{J. Q. Yan}
\affiliation{Ames Laboratory and Department of Physics and Astronomy, Iowa State University, Ames, IA 50011, USA}
\author{E. Mun}
\altaffiliation[Current address: ]{MPA-CMMS, Los Alamos National Laboratory, Los Alamos, NM, 87545}
\affiliation{Ames Laboratory and Department of Physics and Astronomy, Iowa State University, Ames, IA 50011, USA}
\author{P. C. Canfield}
\affiliation{Ames Laboratory and Department of Physics and Astronomy, Iowa State University, Ames, IA 50011, USA}

\date{\today}

\begin{abstract}
Single crystals of Ba(Fe$_{1-x}$Mn$_x$)$_2$As$_2$, $0<x<0.148$, have been grown and characterized by structural, magnetic, electrical transport and thermopower measurements. Although growths of single crystals of Ba(Fe$_{1-x}$Mn$_x$)$_2$As$_2$ for the full $0\leq x\leq 1$ range were made, we find evidence for phase separation (associated with some form of immiscibility) starting for $x>0.1-0.2$. Our measurements show that whereas the structural/magnetic phase transition found in pure BaFe$_2$As$_2$ at 134~K is initially suppressed by Mn substitution, superconductivity is not observed at any substitution level. Although the effect of hydrostatic pressure up to 20~kbar in the parent BaFe$_2$As$_2$ compound is to suppress the structural/magnetic transition at the approximate rate of 0.9~K/kbar, the effects of pressure and Mn substitution in the $x=0.102$ compound are not cumulative. Phase diagrams of transition temperature versus substitution concentration, $x$, based on electrical transport, magnetization and thermopower measurements have been constructed and compared to those of the Ba(Fe$_{1-x}$TM$_x$)$_2$As$_2$ (TM=Co and Cr) series.
\end{abstract}


\maketitle

\section{introduction}
The discovery of superconductivity (SC) in LaFeAs(O$_{1-x}$F$_x$) (Ref.~\onlinecite{kamihara}) and (Ba$_{1-x}$K$_x$)Fe$_2$As$_2$ (Ref.~\onlinecite{rotter:107006}) in 2008 led to extensive interest in these families of FeAs based compounds.  $T_c$ has risen as high as 56~K for F-substituted RFeAsO (R=rare earth, Ref.~\onlinecite{zhi-an}) and as high as 38~K in K- and Na-substituted AEFe$_2$As$_2$ systems (AE=Ba, Sr, Ca).\cite{rotter:107006}  Superconductivity was also found in Co substituted AEFe$_2$As$_2$ (Ref.~\onlinecite{sefat:117004}) and RFeAsO.\cite{sefat:104505}  More recently, superconductivity has been found in other
BaFe$_2$As$_2$ systems (Refs.~\onlinecite{thaler,sharma,wang,saha,tillman,chu,ning,fang,canfield}), as well as SrFe$_2$As$_2$ and CaFe$_2$As$_2$, where the substitution is a $3d$, $4d$ or $5d$ transition metal.  Although the "electron-doped" AEFe$_2$As$_2$ systems have lower $T_c$ values than the "hole-doped" ones (Refs.~\onlinecite{tillman,chu,ning,fang,canfield}), they have been studied extensively because substitution is more homogeneous in these systems and single crystals can be more easily and reproduceably grown.  In order to understand the conditions for superconductivity in these systems, temperature versus substitution and/or doping level phase diagrams must first be constructed.  Detailed studies have been made for transition-metal~(TM) substituted BaFe$_2$As$_2$ (TM=Co, Ni, Cu, Ru, Rh, Pd, Pt, Ir).\cite{tillman,chu,ning,fang,lester,pratt,canfield,thaler,CanfieldOverview}  For Co, Ni, Cu, Rh and Pd, temperature vs substitution concentration, $x$, and temperature vs electron count, $e$, phase diagrams show similar properties, with the temperature of the structural/magnetic transition, $T_s/T_m$, seen in the parent compound being suppressed and separated in a similar manner with $x$ and $T_c$ evolving in a similar manner with $e$, especially on the overdoped side of the superconducting dome.\cite{thaler,tillman,canfield,CanfieldOverview}

In contrast both with electron doping on the Fe site and hole doping on the AE site, attempts to hole dope on the Fe site do not induce superconductivity (Refs.~\onlinecite{sefatCr,MartyCr,budkoCr,kimMn}), so by comparing the properties of hole and electron doped compounds it may be possible to make inferences about the requirements for (and possibly even the mechanism of) superconductivity. 
Motivated by the similarity of the Co and Ni substituted phase diagrams (Ref.~\onlinecite{thaler,canfield,CanfieldOverview}) and the quite different behavior
of BaFe$_2$As$_2$ upon Cr substitution
(Ref.~\onlinecite{sefatCr,MartyCr,budkoCr}), we have studied the effect of Mn substitution on the Fe site.

\section{experimental methods}
Single crystals of Ba(Fe$_{1-x}$Mn$_x$)$_2$As$_2$ were grown out of self flux using conventional high-temperature solution growth techniques.\cite{tillman,canfield,fisk,euroschool}  FeAs was synthesized in the manner described by Ni, et al.\cite{tillman} and MnAs was synthesized by a similar process.  Small Ba chunks and FeAs/MnAs powder were mixed together. A ratio of Ba:TMAs=1:4 was used.  The mixture was then placed in an alumina crucible with a "catch" crucible filled with silica wool placed on top.\cite{fisk,euroschool} Both crucibles were sealed in a silica tube under a 1/6, partial atmosphere, of Ar gas.  The sealed tube was gradually heated to 1180$^{\circ}$C over 19 hours, held at 1180$^{\circ}$C for 8-12 hours, and then cooled over 40-65 hours to a final temperature of 1000$^{\circ}$C. Once the furnace reached the final temperature, the excess FeAs/MnAs liquid was decanted (Ref.~\onlinecite{fisk,euroschool}), leaving the plate-like single crystals exposed for easy removal from the growth crucible.

Powder x-ray diffraction measurements, with a Si standard, were performed using a Rigaku Miniflex diffractometer with Cu $K\alpha$ radiation at room temperature. Diffraction patterns were taken on ground single crystals from each batch. Only very small FeAs impurity peaks (associated with residual flux) were found as a secondary phase. The unit cell parameters were refined using "Rietica" software.  Elemental analysis of single crystal samples was used to determine the actual percentage of the substitution in the lattice as opposed to the nominal substitution level.  This was performed using wavelength dispersive x-ray spectroscopy (WDS) in the electron probe microanalyzer of a JEOL JXA-8200 electron-microprobe.  Magnetization data were collected in a Quantum Design (QD) Magnetic Properties Measurement System (MPMS).  Temperature-dependent AC electrical resistivity data (f=16~Hz, I=3~mA) were collected using either a QD MPMS with a LR700 resistance bridge or a QD Physical Properties Measurement System (PPMS). Electrical contact was made to the sample using Epotek H20E silver epoxy to attach Pt wires in a four-probe configuration. Thermoelectric power (TEP) measurements were performed using a dc, alternating temperature gradient (two heaters and two thermometers) technique (Ref.~\onlinecite{edmtep}) using a Quantum Design PPMS to provide the thermal environment. The samples were bridged between two Cernox (CX-1050) temperature sensors, using DuPont 4929N silver paste to ensure good thermal and electrical contact.

Measurements of dc electrical resistivity under hydrostatic pressures up to $\sim$20~kbar were carried out in a Be-Cu piston-cylinder cell with a center core of tungsten carbide. A feedthrough containing the sample along with manganin and Pb manometers was inserted in a Teflon cup filled with a 40:60 mixture of light mineral oil and n-pentane, which served as the pressure-transmitting medium. Pressure was applied and locked in at ambient temperature with a hydraulic press, using the manganin manometer as a reference. The pressure in this type of cell is known to decrease with decreasing temperature, before stabilizing below $\sim 90$~K.\cite{thompson} The pressure at low temperatures was determined from the superconducting transition temperature of the Pb manometer, and pressure values between ambient temperature and 90~K were estimated by linear interpolation. The pressure cell was placed in the PPMS, and the temperature of the sample was measured with a calibrated Cernox sensor attached to the body of the cell. In order to minimize the temperature lag between sample and sensor, the temperature sweep rate was kept below 0.35~K/min.

\section{results}
\begin{figure}[h]
	\begin{center}
		\includegraphics[width=.9\linewidth]{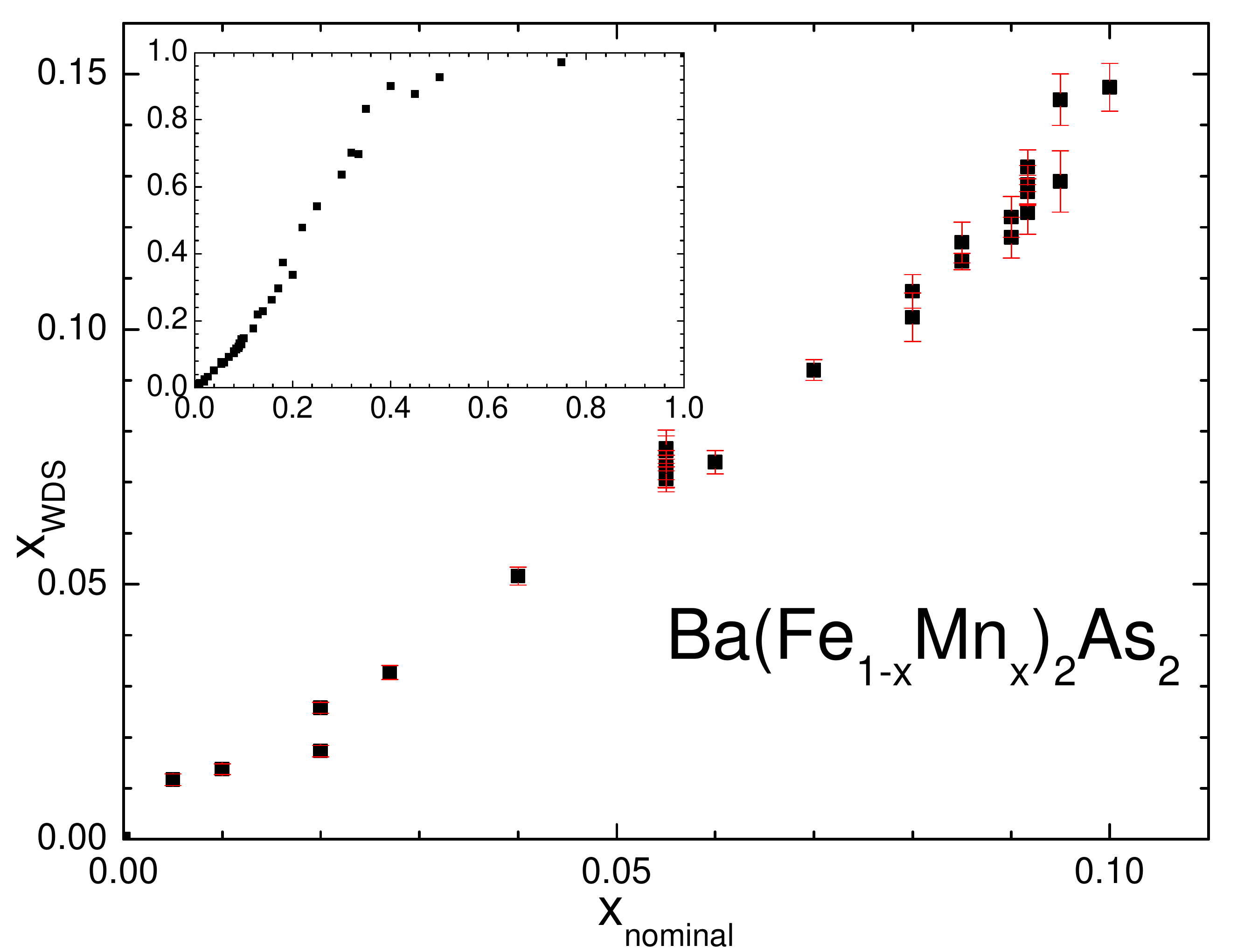}
	\end{center}
	\caption{Experimentally determined Mn concentration, $x_{WDS}$, vs nominal Mn concentration.  Error bars are $\pm2\sigma$ (values from Table~\ref{table:WDSdata}). Inset shows full substitution range. (Color online)}
	\label{fig:rvn}
\end{figure}

WDS was performed in order to determive actual rather than nominal substitution concentration. Each WDS measurement provides composition information about sample volumes of $\sim1$~$\mu m$ diameter. WDS measurement data are summarized in Table~\ref{table:WDSdata} and Fig.~\ref{fig:rvn}.  For each batch, up to 5 crystalline surfaces were examined.  The table shows the number of points measured, the average $x$ value, and two times the standard deviation of the $x$ values measured.  Unless otherwise noted, all $x$ values given in this paper are the average $x_{WDS}$ values determined by WDS, not the nominal $x$ substitution. As shown, there is a clear tendency for $x>x_{\text{nominal}}$. For $x<0.09$ the slope of the $x(x_{nominal})$ line is $\frac{\Delta x}{\Delta x_{nominal}}\approx1.3$. For $x>0.15$, this trend becomes dramatically stronger: $x$ shoots upward (in comparison to $x_{nominal}$) and hints at possible phase stability or phase separation problems (see inset to Fig.~\ref{fig:rvn}).

\onecolumngrid
\begin{center}
\begin{table}[h]
	\begin{tabular}{c|c|c|c|c|c|c|c|c|c|c|c|c}
	\hline\hline
	\multicolumn{12}{c}{Ba(Fe$_{1-x}$Mn$_x$)$_2$As$_2$}\\
	\hline
N	&	15	&	12	&	12	&	12	&	13	&	9	&	12	&	13	&	12	&	12	&	14	\\
\hline
$x_{nominal}$	&	0.01	&	0.02	&	0.02	&	0.027	&	0.04	&	0.06	&	0.07	&	0.08	&	0.09	&	0.095	&	0.1	\\
\hline
$x_{WDS}$	&	0.014 &	0.017	&	0.026	&	0.033	&	0.052	&	0.074	&	0.092	&	0.102	&	0.118	&	0.129	&	0.148	\\
\hline
$2\sigma$	&	0.001	&	0.001	&	0.001	&	0.001	&	0.002	&	0.002	&	0.002	&	0.005	&	0.004	&	0.006	&	0.005	\\
\hline
\hline
	\end{tabular}
	\caption{WDS data for Ba(Fe$_{1-x}$Mn$_x$)$_2$As$_2$, $x<0.15$.  N is the number of points measured in each batch, $x_{WDS}$ is the average $x$ value for that batch, and $2\sigma$ is twice the standard deviation of the N values measured.}
	\label{table:WDSdata}
\end{table}
\end{center}
\twocolumngrid

\begin{figure}[h]
	\begin{center}
		\includegraphics[width=.9\linewidth]{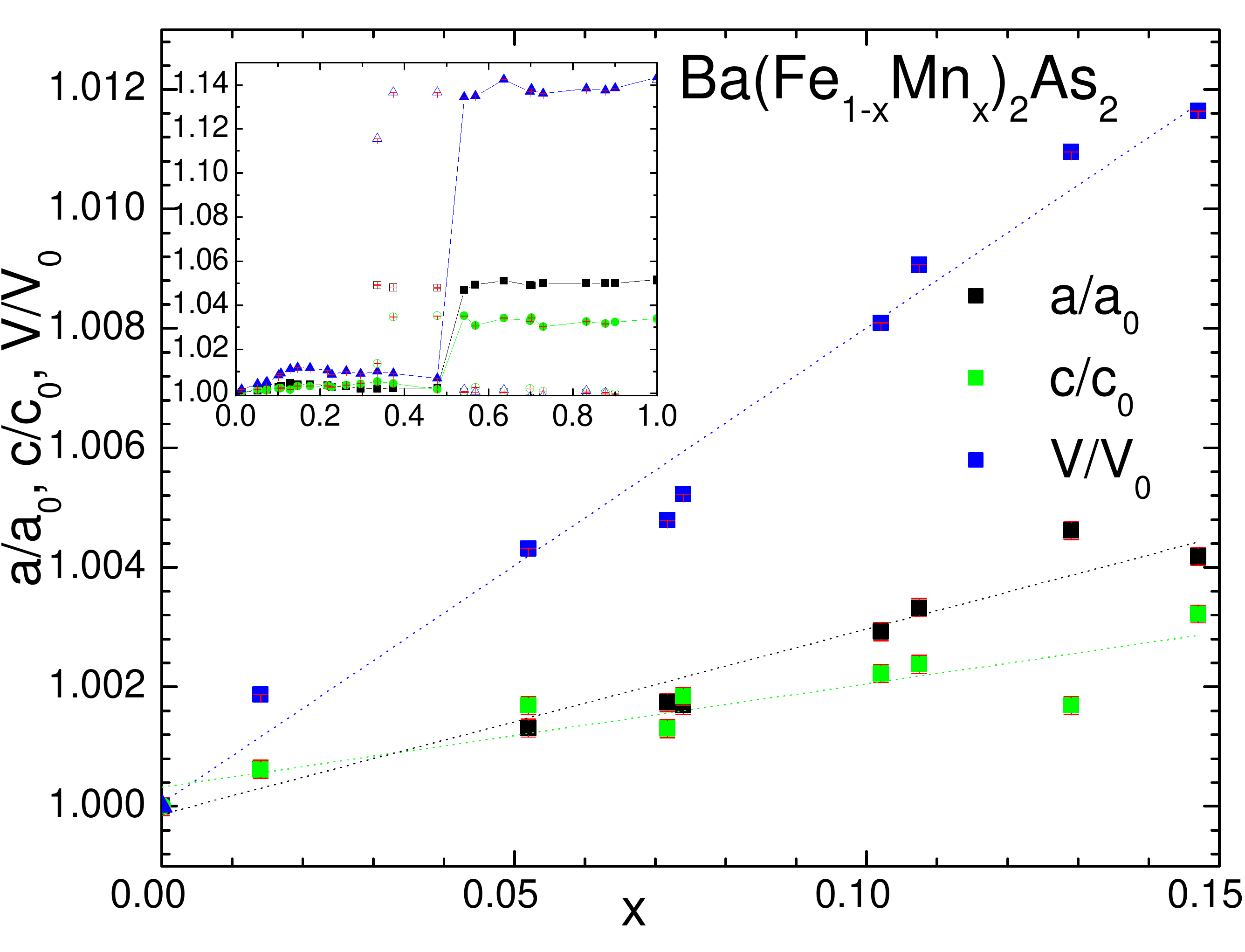}
	\end{center}
	\caption{Unit cell parameters, $a$ and $c$, as well as volume, $V$, normalized to those of the parent compound BaFe$_2$As$_2$, for which $a_0=3.96\text{~\AA}$, $c_0=13.0\text{~\AA}$ and $V_0=204\text{~\AA}^3$.  The open symbols at $x=0$ are previously published data for the parent compound. Dashed lines are guides to the eye. Inset: full range $0<x<1$. Open symbols are minority phase lattice parameters.
(Color online)}
	\label{fig:LPplot}
\end{figure}

Powder x-ray diffraction data is consistent with Ba(Fe$_{1-x}$Mn$_x$)$_2$As$_2$ forming in the I4/mmm, ThCr$_2$Si$_2$ structure. Rietvelt refinement of powder x-ray patterns using Si as a standard allows lattice parameters to be determined, from which unit cell volumes can be calculated. Normalized lattice parameters and unit cell volumes are displayed in Fig.~\ref{fig:LPplot}. There is a monotonic increase in the $a$ and $c$ lattice parameters and unit cell volume as $x$ increases from zero up through $x\sim0.1$.
This is consistent with the partial substitution of Fe with Mn.\cite{thaler,sefatCr} The saturation of $a$ and $c$ for $x$ near $0.15$ is a departure from Vegard's rule, and it suggests a limit of phase stability.

\begin{figure}[h]
	\begin{center}
		\includegraphics[width=.9\linewidth]{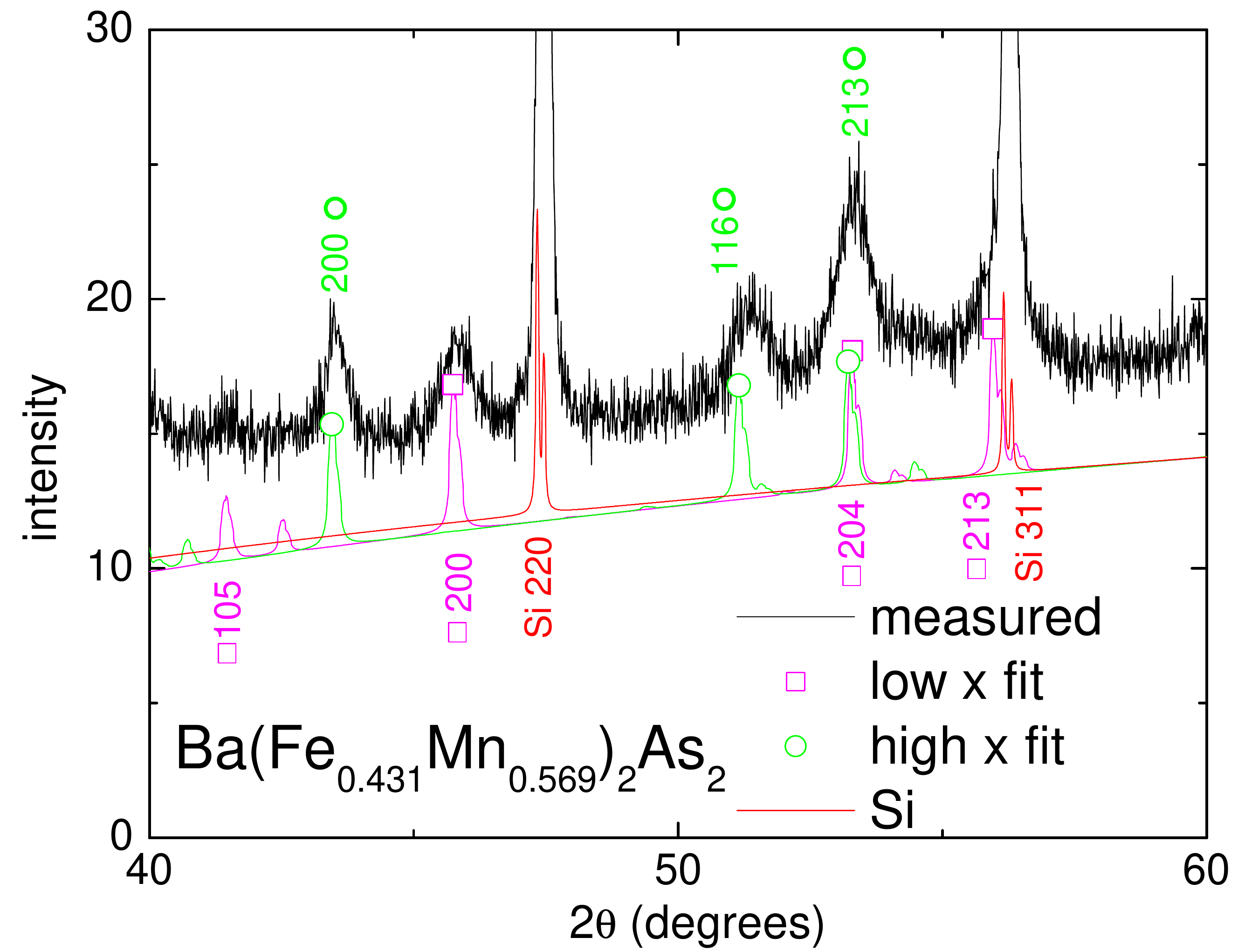}
	\end{center}
	\caption{Powder x-ray pattern for $x=0.569$, showing both high and low concentration fits. The strong peaks near 47$^\circ$ and 56$^\circ$ are from the Si standard. The 56$^\circ$ peak is near to the hkl=213 peak expected from the lower concentration fit. (Color online)}
	\label{fig:dirtyxray}
\end{figure}

As shown in the inset of Fig.~\ref{fig:LPplot}, there is a large jump in the size of a- and c-lattice parameters near $x=0.5$, inferred from the fits of the diffractograms.
Between $x\approx0.336$ and $x\approx 0.6$, two phases can be detected. Below $x=0.5$, the majority phase has lattice parameters closer to the $x=0$ values and the minority phase lattice parameters are closer to the $x=1$ values. Above $x\sim0.5$, the majority is near $x=1$ and the minority is near $x=0$.
The presence of these two, closely related phases can be seen for a representative diffractogram in Fig.~\ref{fig:dirtyxray}: neither phase can fit all of the peaks alone, but together they adequately describe all of the major features seen.
The ratio of the two 200 peaks is close to unity for $x\sim0.55$. On either side of $x\sim0.55$, the scattering power of the minority phases decreases rapidly.

These results are consistent with a clear phase separation as described in detail in Ref.~\onlinecite{pandey}. The lattice parameters of BaMn$_2$As$_2$ are $a$~$=$~$4.1686$~\AA~and $c$~$=$~$13.473$~\AA. This is a much larger deviation from BaFe$_2$As$_2$ ($a$~$=$~$3.9653$~\AA~and $c$~$=$~$13.036$~\AA, Ref.~\onlinecite{tillman}) than either BaCo$_2$As$_2$ ($a$~$=$~$3.9537$~\AA~and $c$~$=$~$12.6524$~\AA, Ref.~\onlinecite{SefatCo}) or BaCr$_2$As$_2$ ($a$~$=$~$3.9678$~\AA~and $c$~$=$~$13.632$~\AA, Ref.~\onlinecite{sefatCr}). This is the likely source of the immiscibility at high $x$ in Mn substitution. The source of this large change in lattice parameters is not fully clear: the free atomic radii of Cr, Mn, Fe and Co are all very close (Ref.~\onlinecite{thaler,sefatCr}), with Co deviating the most from Fe. The ionic radii of the four transition metals are also similar (depending on the charge and spin state), with Cr deviating the most from Fe.\cite{ionicradii} Thus, steric effects alone are not enough to account for the large shift in lattice parameters when comparing BaMn$_2$As$_2$ with its transition metal neighbors. It is likely that the strong magnetism of the Mn atoms is at least partly responsible for the difference,
most probably via high spin/low spin effects.\cite{pandey}

For these reasons, we limit our investigation of thermodynamic and transport properties of this series to $x<0.15$, where the parameters shown in Fig.~\ref{fig:LPplot} still roughly follow Vegard's law. It is important to note, though, that we did not see any such immiscibility manifesting as a dramatic increase in the standard deviation of $x$ (see Table.~\ref{table:WDSdata}). This implies that the spot volume probed by the electron beam of the WDS microprobe ($\sim 1 \mu m^3$) is larger than the characteristic length scale of the phase separated regions.

\begin{figure}[h]
	\begin{center}
		\includegraphics[width=.9\linewidth]{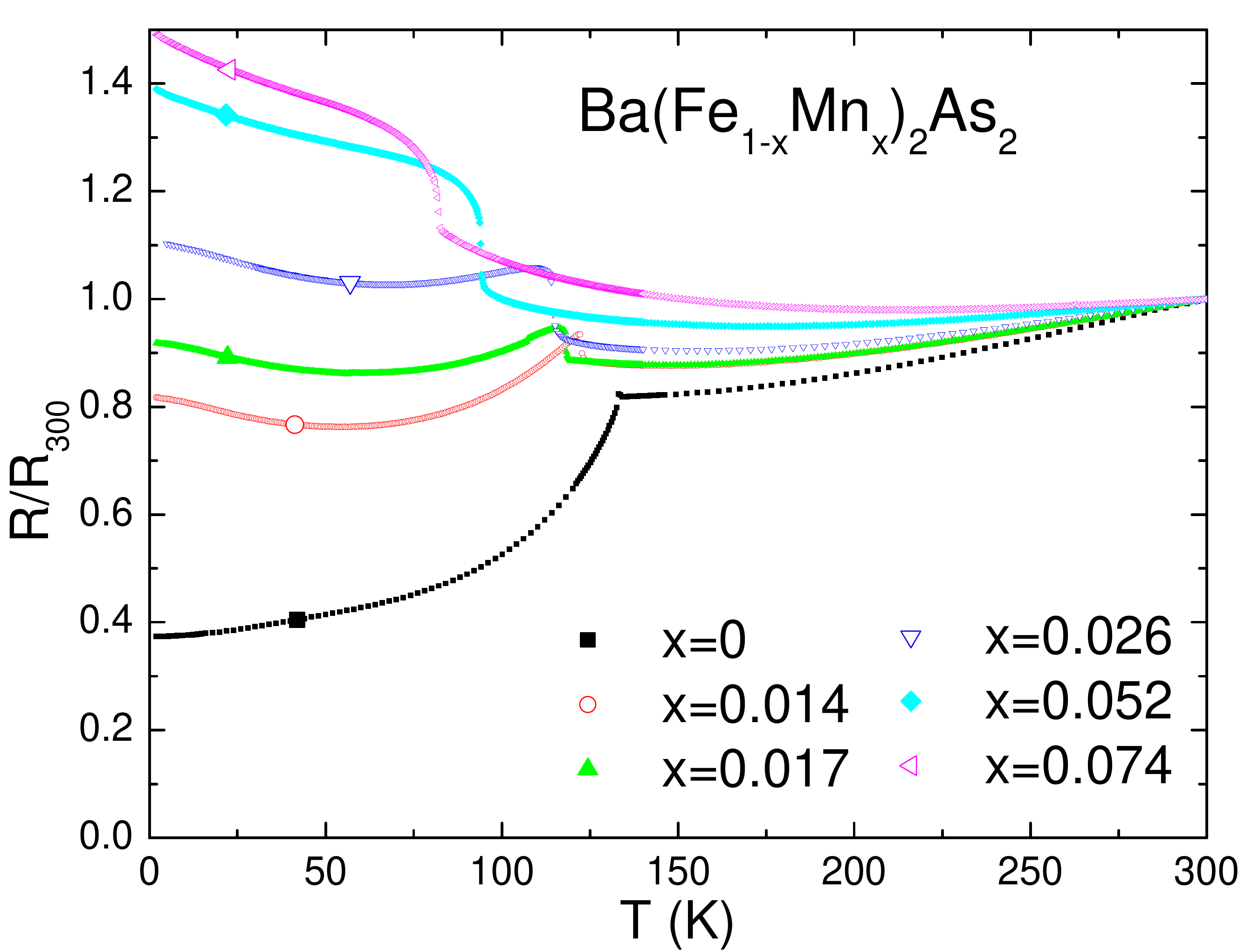}
		\includegraphics[width=.9\linewidth]{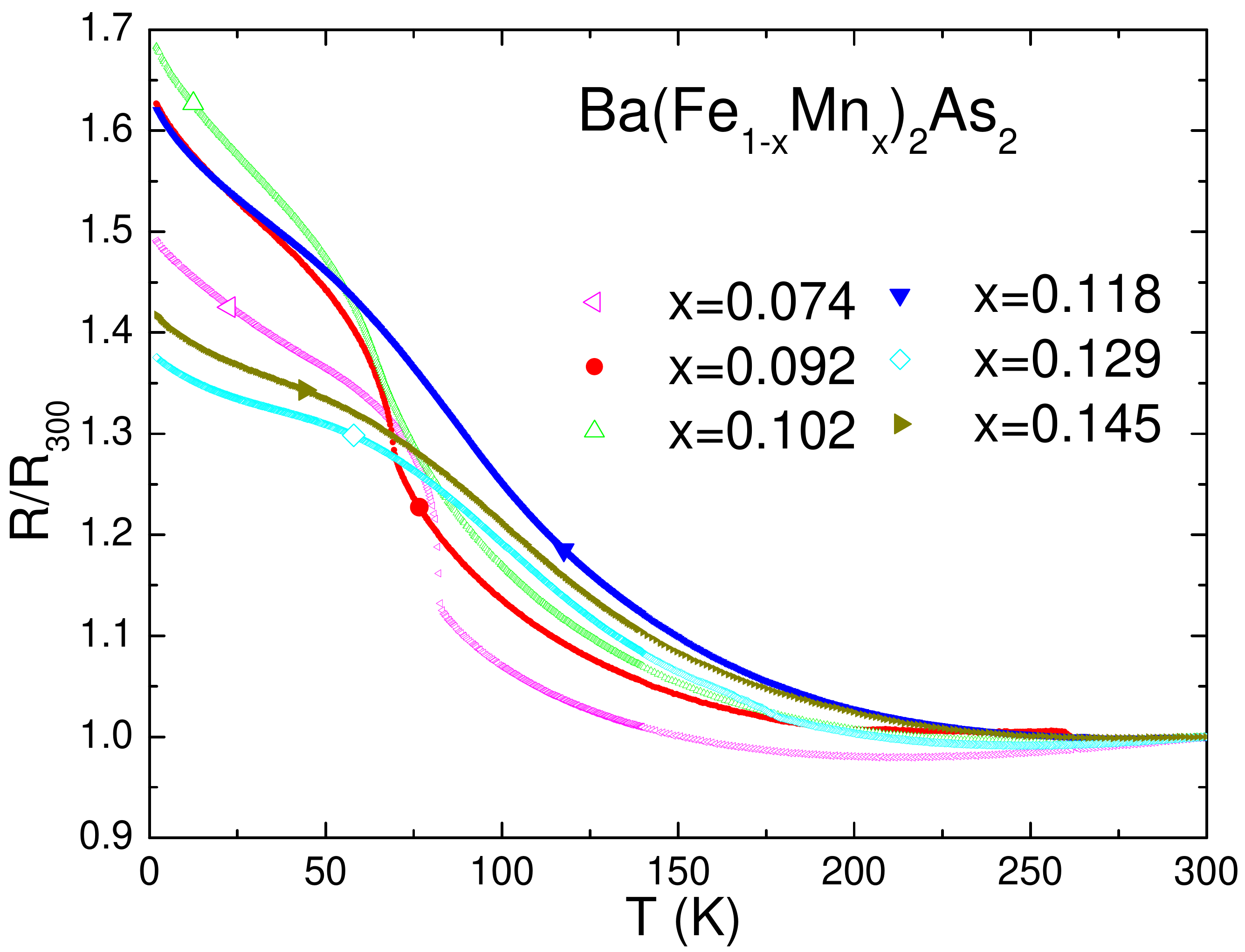}
	\end{center}
	\caption{Temperature dependent resistivity, normalized to the room temperature value, for Ba(Fe$_{1-x}$Mn$_x$)$_2$As$_2$. The $x=0.074$ data are shown in both panels for the sake of comparison. (Color online)}
	\label{fig:RRRdata}
\end{figure}

\begin{figure}[h]
	\begin{center}
		\includegraphics[width=0.9\linewidth]{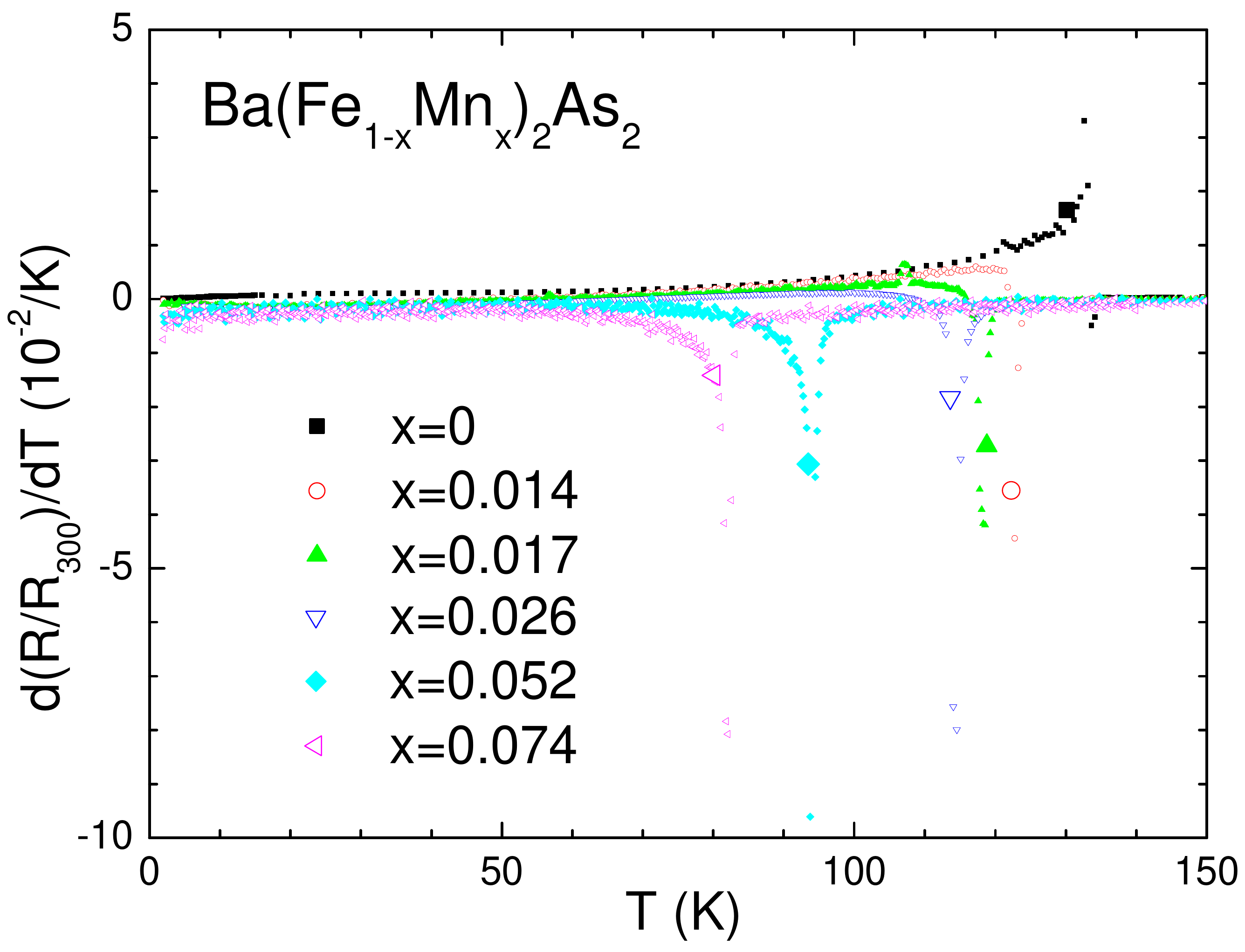}
		\includegraphics[width=0.9\linewidth]{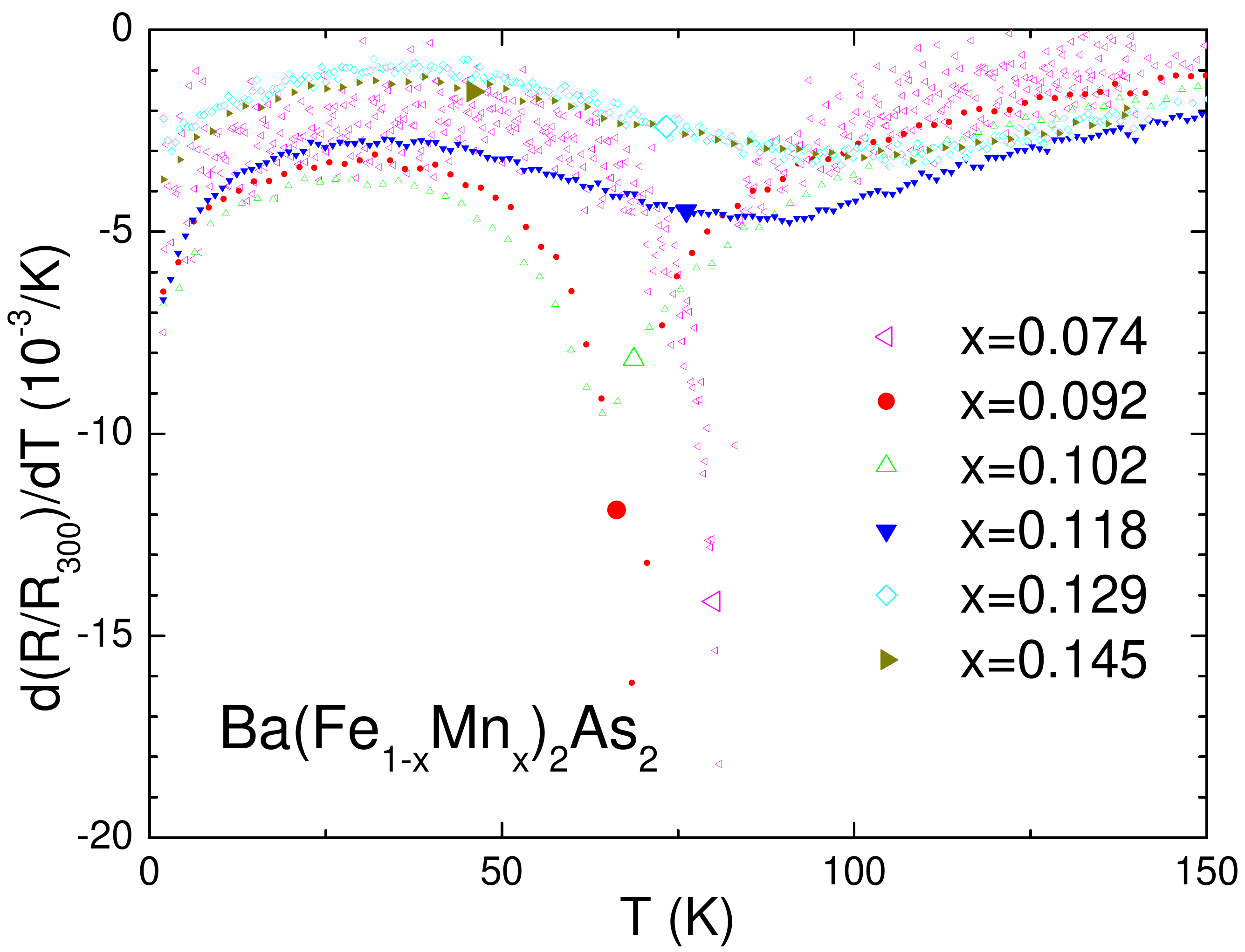}
	\end{center}
	\caption{First derivative of normalized resistivity for Ba(Fe$_{1-x}$Mn$_x$)$_2$As$_2$. The $x=0.074$ data are shown in both panels for the sake of comparison. (Color online)}
	\label{fig:resderiv}
\end{figure}

Fig.~\ref{fig:RRRdata} shows the normalized electrical resistivity data of the Ba(Fe$_{1-x}$Mn$_x$)$_2$As$_2$ series from 2 K to 300 K.  Normalized resistivity is plotted instead of resistivity because of the tendency of these samples to exfoliate or crack.\cite{tillman,tanatar:094507,tanatar:134528}  The anomaly in resistivity at 134 K for pure BaFe$_2$As$_2$ is associated with the structural/magnetic phase transitions (Ref.~\onlinecite{rotter:020503}), and can clearly be seen in the derivative of the resistivity (Ref.~\onlinecite{tillman}), which is shown in Fig.~\ref{fig:resderiv}. As in the case of other TM substitutions (Refs.~\onlinecite{tillman,canfield,thaler,CanfieldOverview}), with Mn substitition the temperature of the resistive anomaly is suppressed and the shape is changed from a sharp decrease to a broadened increase as $x$ increases.  The anomaly becomes extremely broad around $x=0.1$, and is no longer detectable by $x=0.147$. For $T>2K$, superconductivity is not observed for any measured substitution level.


Although this low $x$ anomaly in resistivity
disappears slightly above $x\sim0.1$ (Figs.~\ref{fig:RRRdata}(a)~and~\ref{fig:resderiv}(a)), there is still a feature of note at higher substitution levels (Figs.~\ref{fig:RRRdata}(b)~and~\ref{fig:resderiv}(b)).  For $x>0.1$, a broad minimum in the derivative of the resistance, associated with a maximal, negative $dR/dT$ value, is evident.  The temperature of this minimum gradually increases with substitution level and, as will be discussed below, follows the temperature above which neutron scattering no longer observes magnetic scattering characteristic of long range order.\cite{kimMn}

\begin{figure}[h]
	\begin{center}
		\includegraphics[width=.9\linewidth]{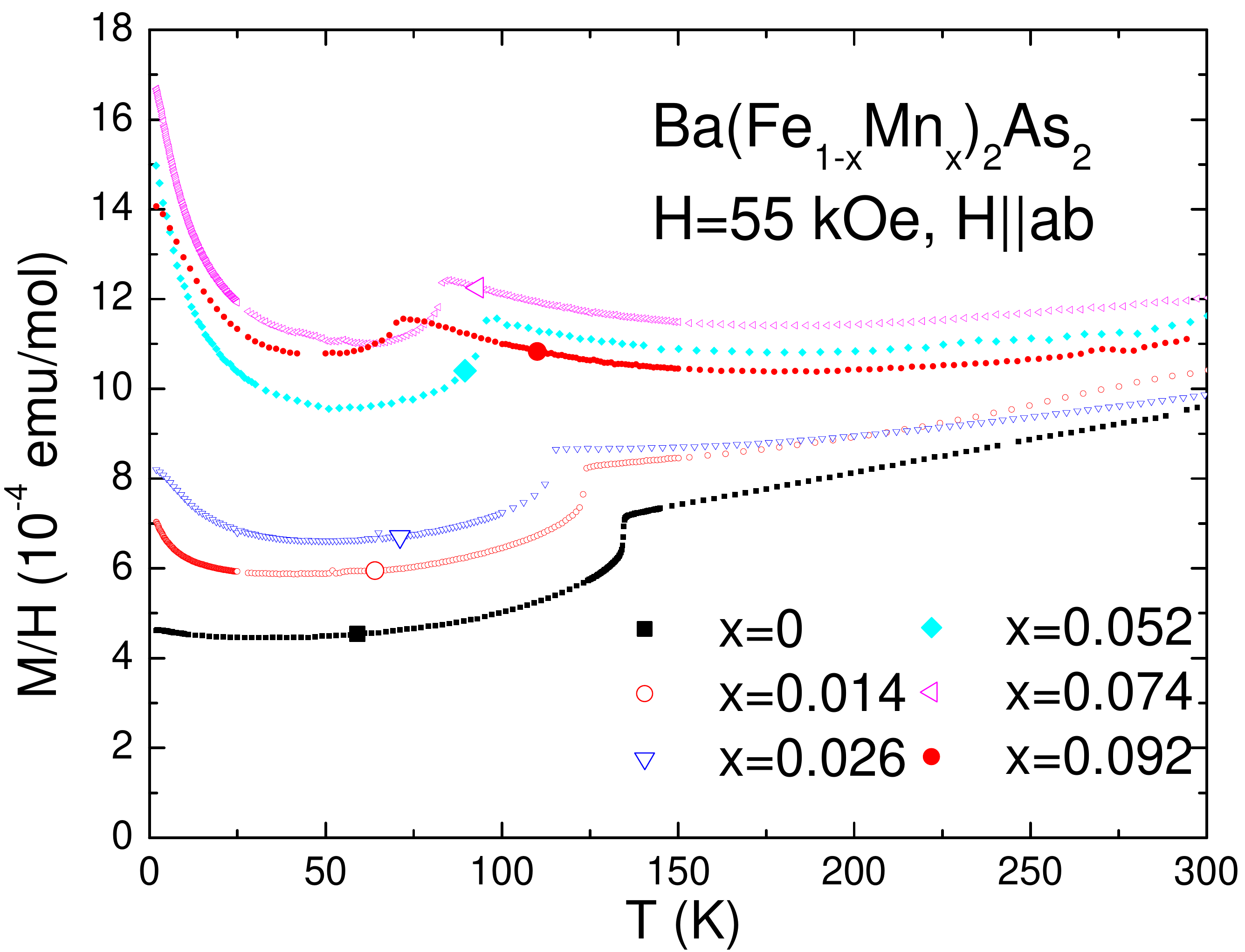}
		\includegraphics[width=.9\linewidth]{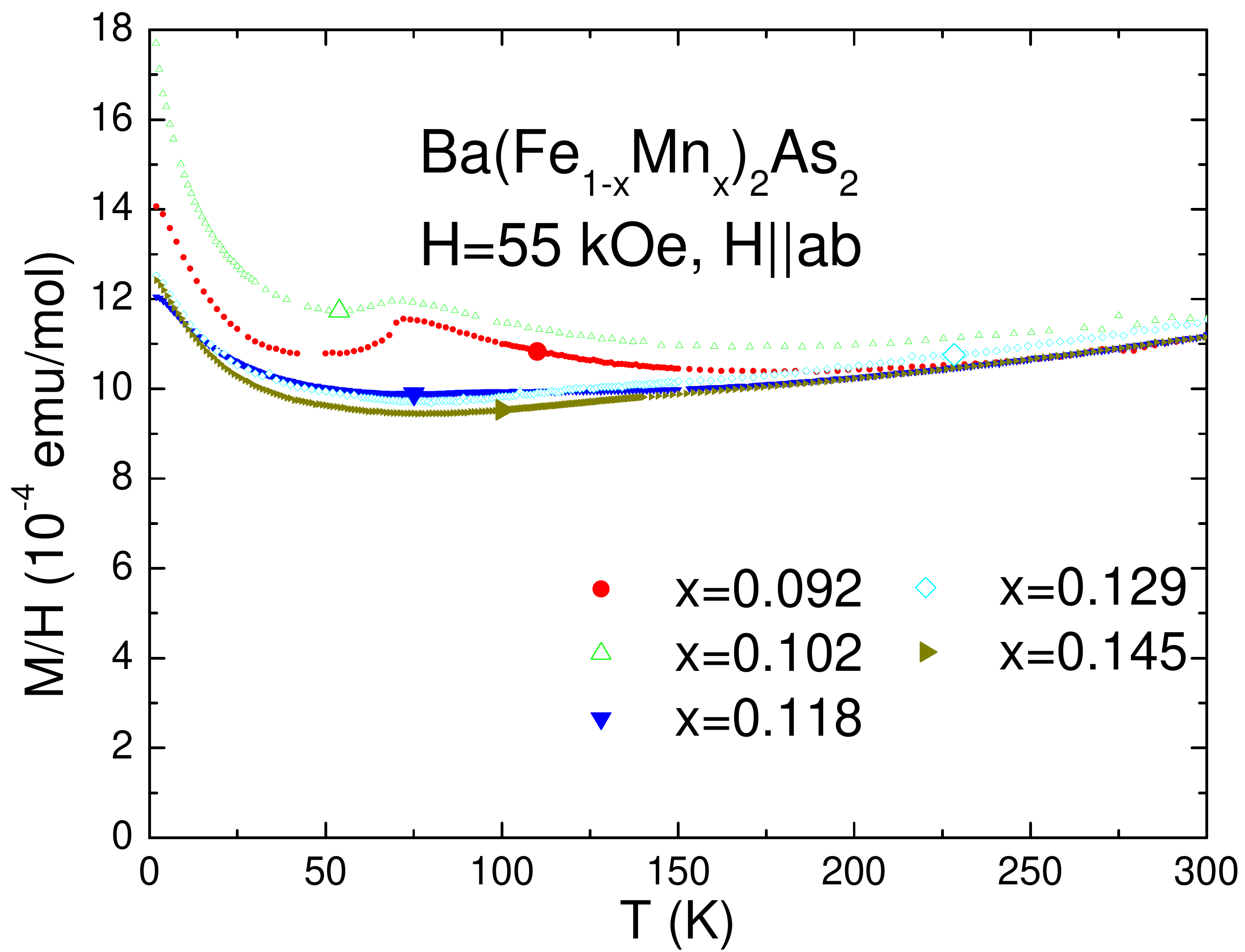}
	\end{center}
	\caption{Temperature dependent magnetization for Ba(Fe$_{1-x}$Mn$_x$)$_2$As$_2$ with $H||ab$.  In all cases, $H=55$~$kOe$. The $x=0.092$ data are shown in both panels for the sake of comparison. (Color online)}
	\label{fig:MTdata}
\end{figure}

\begin{figure}[h]
	\begin{center}
		\includegraphics[width=.9\linewidth]{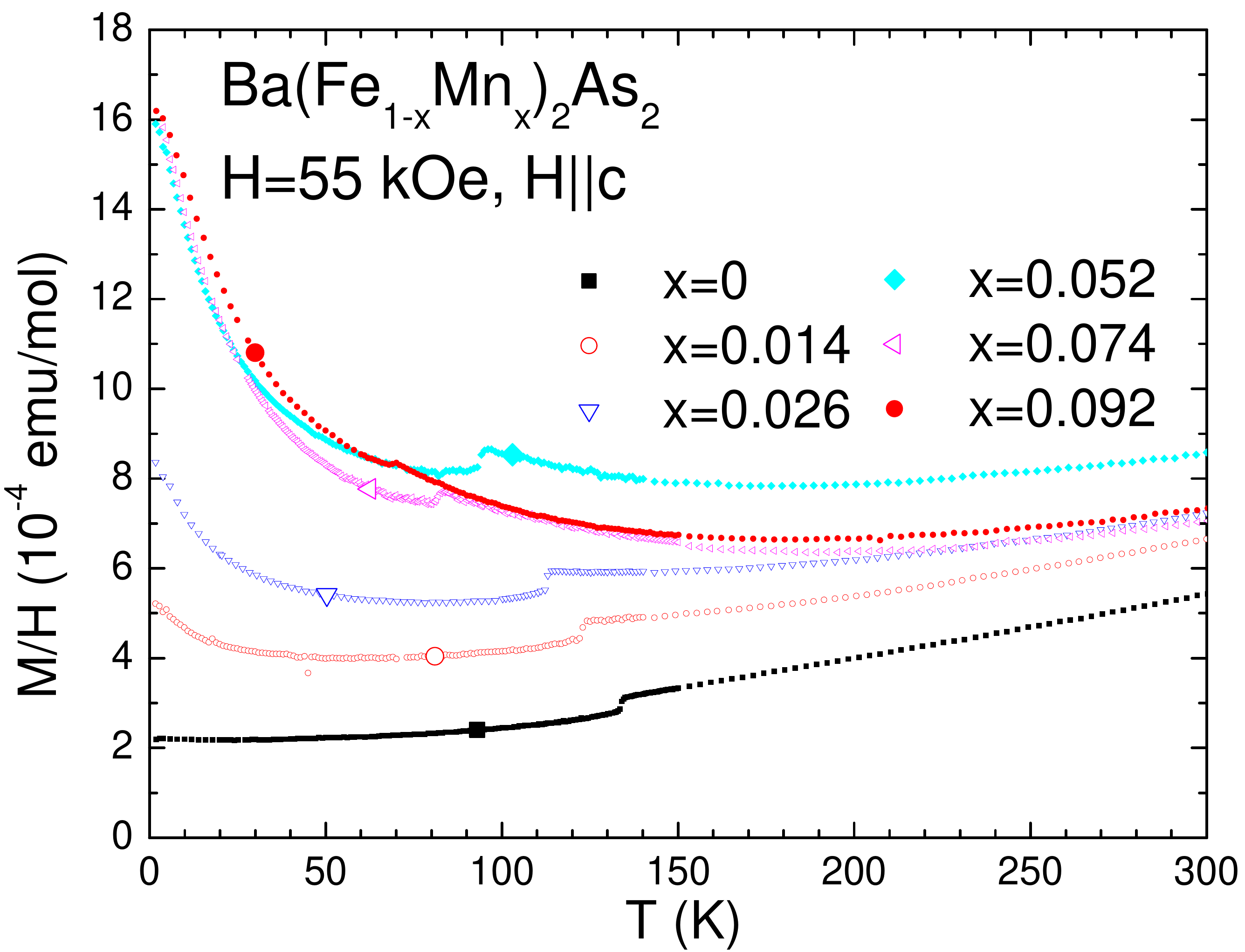}
		\includegraphics[width=.9\linewidth]{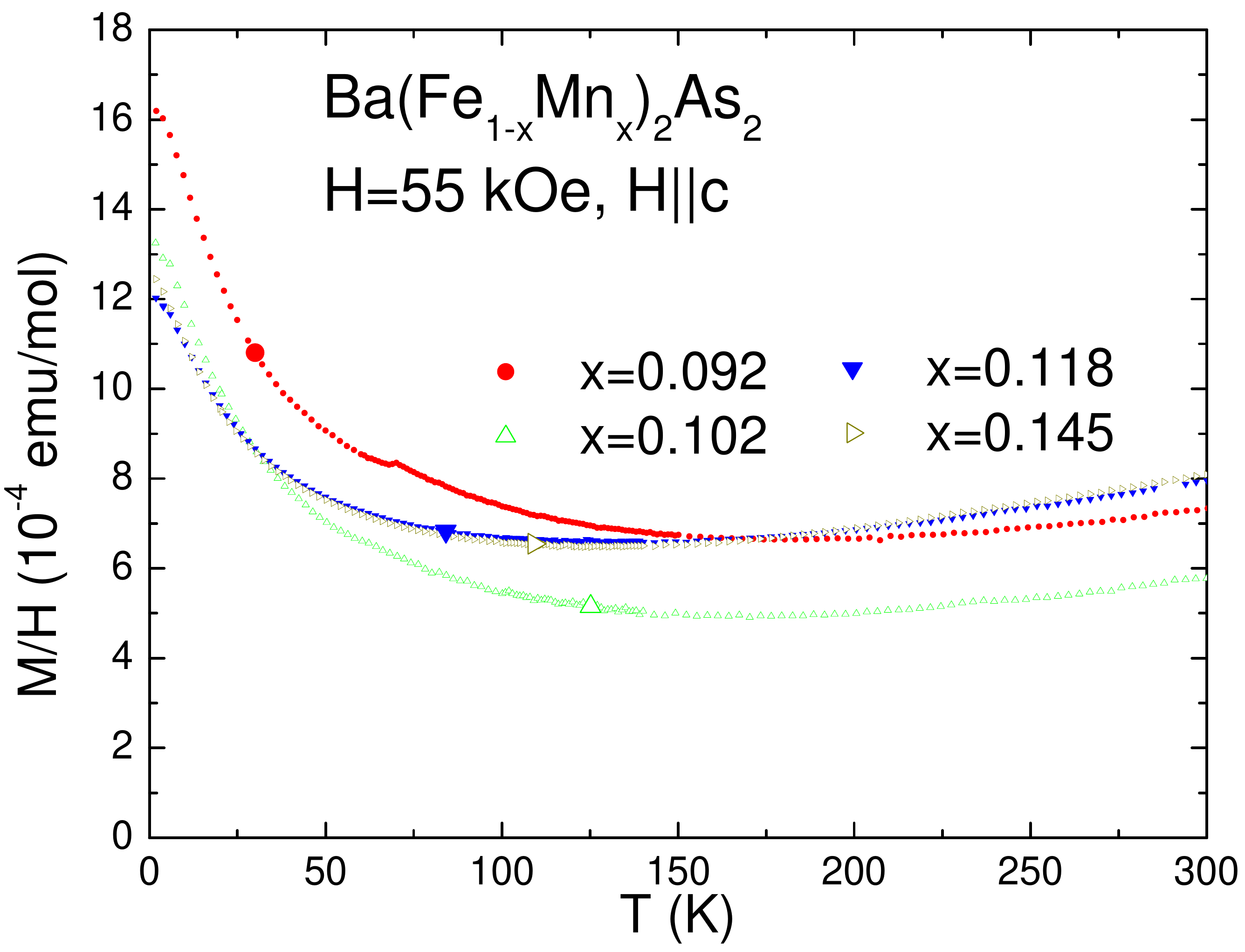}
	\end{center}
	\caption{Temperature dependent magnetization for Ba(Fe$_{1-x}$Mn$_x$)$_2$As$_2$ with $H||c$.  In all cases, $H=55$~$kOe$. The $x=0.092$ data are shown in both panels for the sake of comparison. (Color online)}
	\label{fig:MTH||cdata}
\end{figure}

\begin{figure}[h]
	\begin{center}
		\includegraphics[width=.9\linewidth]{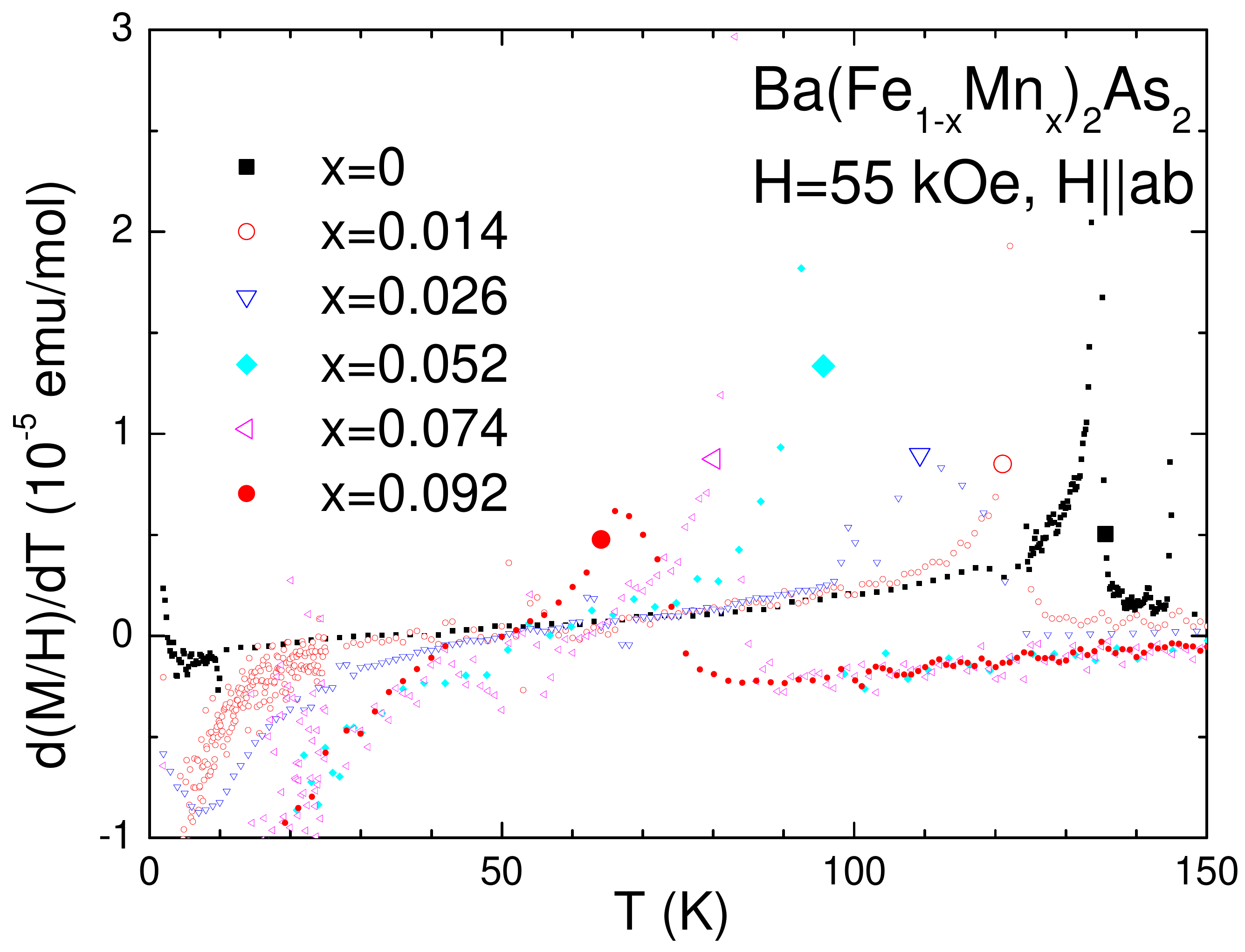}
		\includegraphics[width=.9\linewidth]{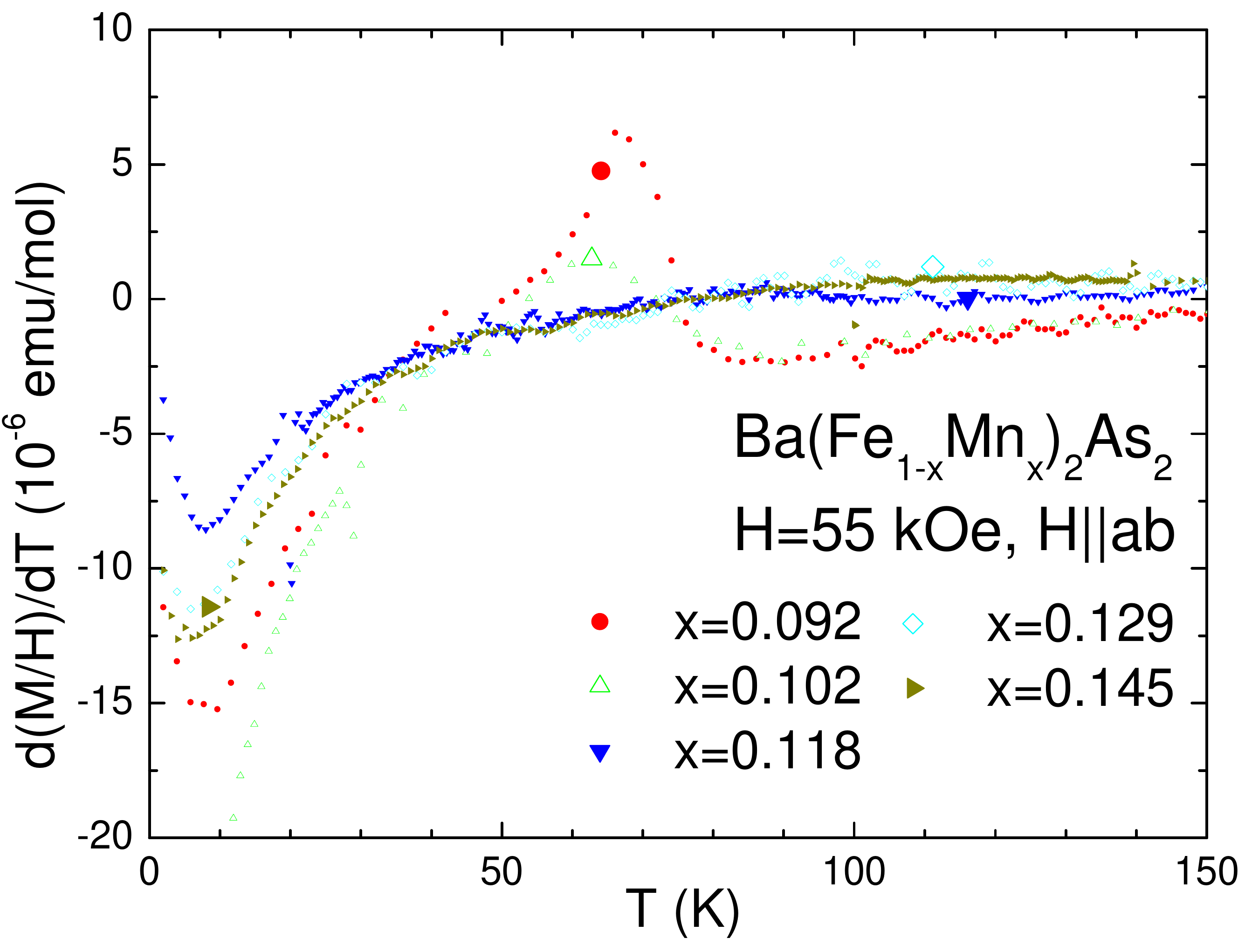}
	\end{center}
	\caption{Derivatives of the temperature dependent magnetization for Ba(Fe$_{1-x}$Mn$_x$)$_2$As$_2$ with $H||ab$. In all cases, $H=55$~$kOe$. The $x=0.092$ data are shown in both panels for the sake of comparison. (Color online)}
	\label{fig:MTderivdata}
\end{figure}

\begin{figure}[h]
	\begin{center}
		\includegraphics[width=.9\linewidth]{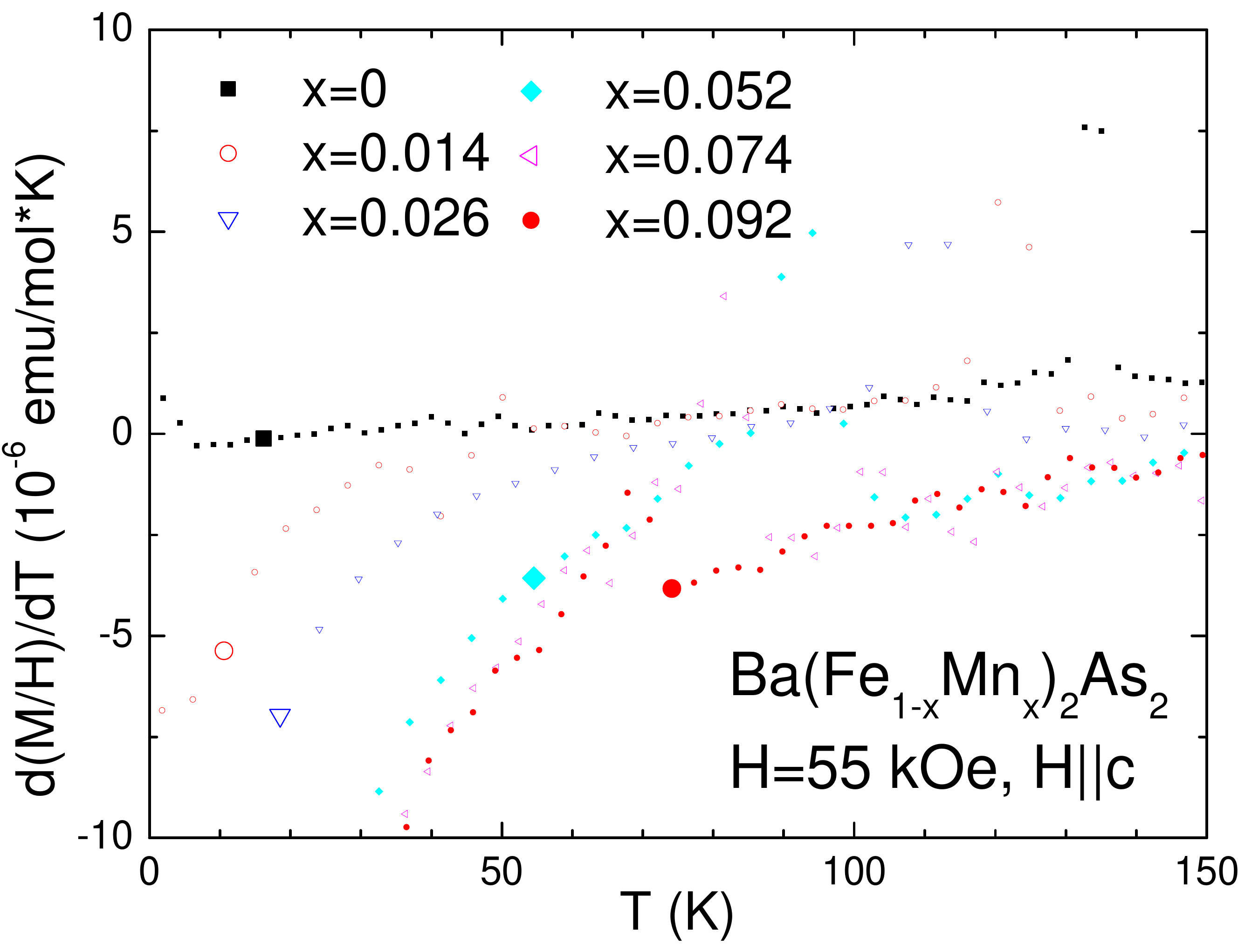}
		\includegraphics[width=.9\linewidth]{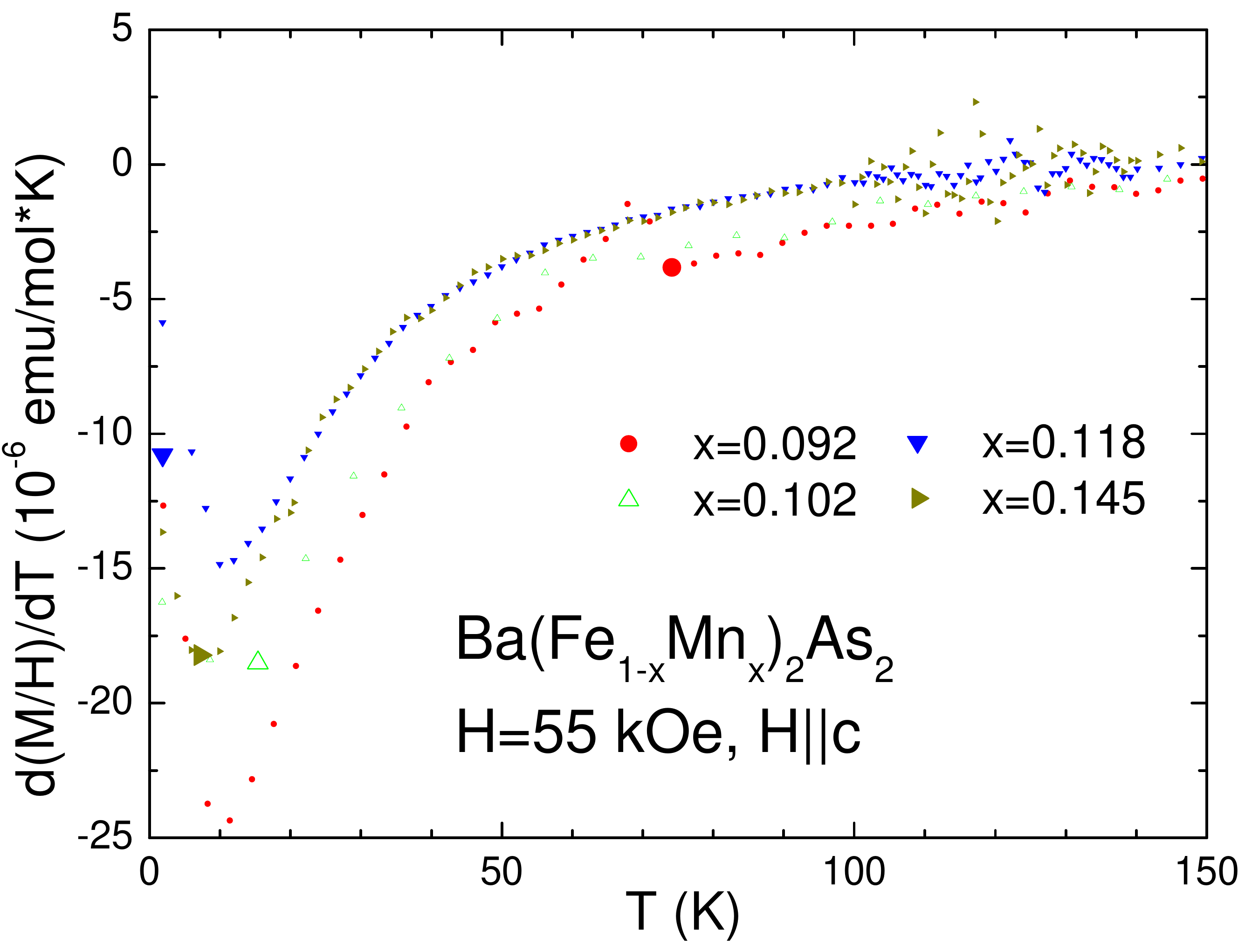}
	\end{center}
	\caption{Derivatives of the temperature dependent magnetization for Ba(Fe$_{1-x}$Mn$_x$)$_2$As$_2$ with $H||c$. In all cases, $H=55$~$kOe$. The $x=0.092$ data are shown in both panels for the sake of comparison. (Color online)}
	\label{fig:MTderivH||cdata}
\end{figure}

Signatures of the structural/magnetic phase transition in the lower substitution region are also seen in magnetization data: Figs.~\ref{fig:MTdata} and \ref{fig:MTH||cdata} show temperature dependent magnetization data of the Ba(Fe$_{1-x}$Mn$_x$)$_2$As$_2$ series from 2~K to 300~K for both directions of the applied field and Figs.~\ref{fig:MTderivdata} and \ref{fig:MTderivH||cdata} show the derivatives of the temperature dependent magnetization data.
For both directions of the applied field, a sharp feature can be seen at the structural/magnetic transition temperature for $x\leq0.092$. For $x=0.102$ the feature has broadened and become harder to resolve, and for $x>0.102$ no feature that can be easily associated with a transition can be found in the $M(T)$ or $dM(T)/dT$ data. It is also at this concentration that the general trend of $M(T)$ increasing with increasing $x$ ends.

\begin{figure}[h]
	\begin{center}
		\includegraphics[width=.9\linewidth]{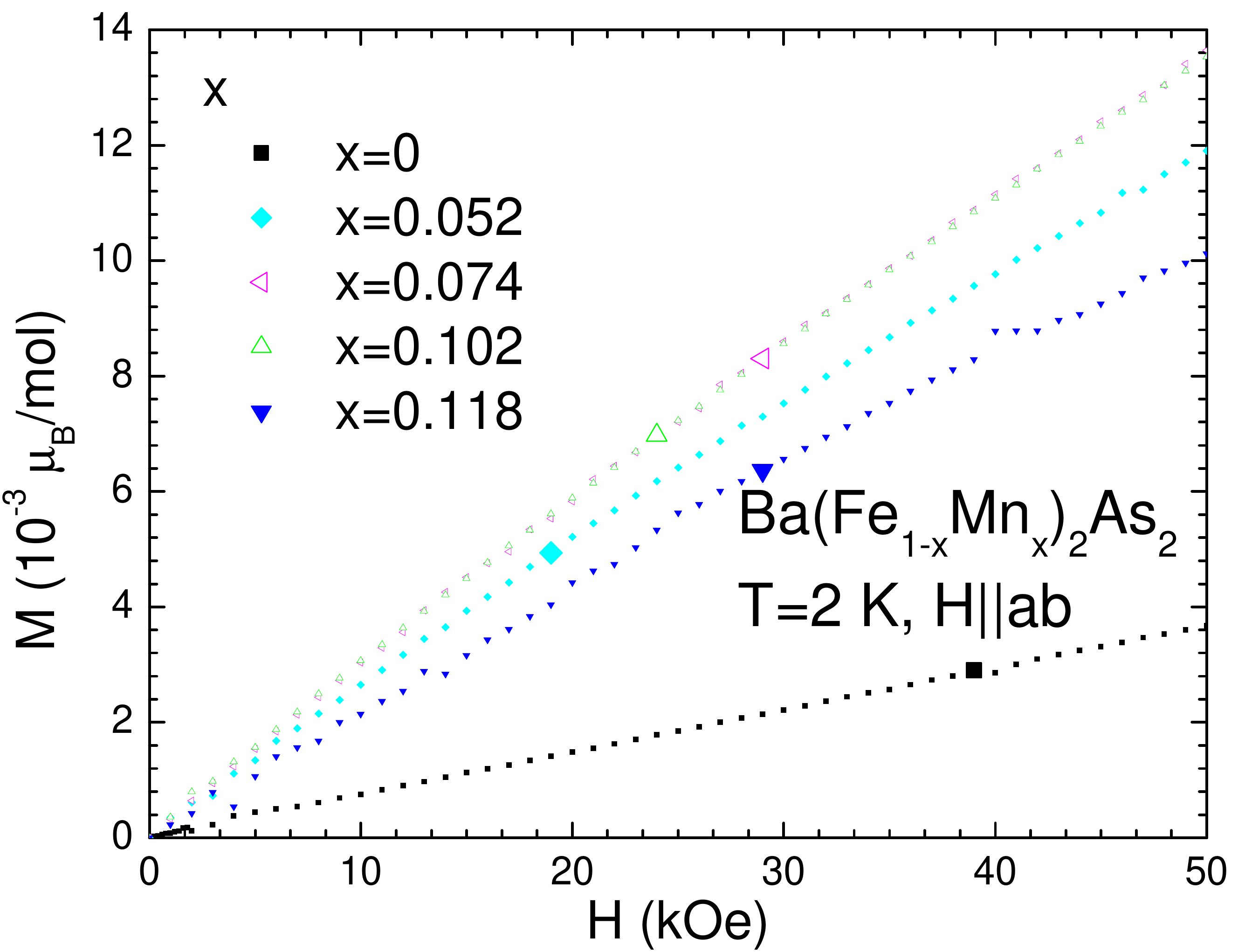}
	\end{center}
	\caption{Field dependent magnetization for Ba(Fe$_{1-x}$Mn$_x$)$_2$As$_2$, $H||ab$.  In all cases, $T=2K$. (Color online)}
	\label{fig:MHdata}
\end{figure}

Fig.~\ref{fig:MHdata} shows the field dependent magnetization ($H||ab$) of the Ba(Fe$_{1-x}$Mn$_x$)$_2$As$_2$ series at 2~K. In the substitution range we have explored, there is no evidence of a ferromagnetic component to the magnetization, and only slight non-linearity that could be associated with Brillouin saturation of paramagnetic local moments. This suggests that if there is a little residual flux present in these samples, it does not contain MnAs since MnAs is a ferromagnet with $T_C\approx318$~K.\cite{singh}
Although it is tempting to try to associate the low temperature upturn in magnetization, seen most clearly for intermediate $x$ values, with a Curie tail coming from paramagnetic impurities, this is difficult.
Since the size of this upturn-like feature diminishes as $x$ increases for higher substitution levels, it is hard to simply associate with increased Mn levels.



\begin{figure}[h]
	\begin{center}
		\includegraphics[width=.9\linewidth]{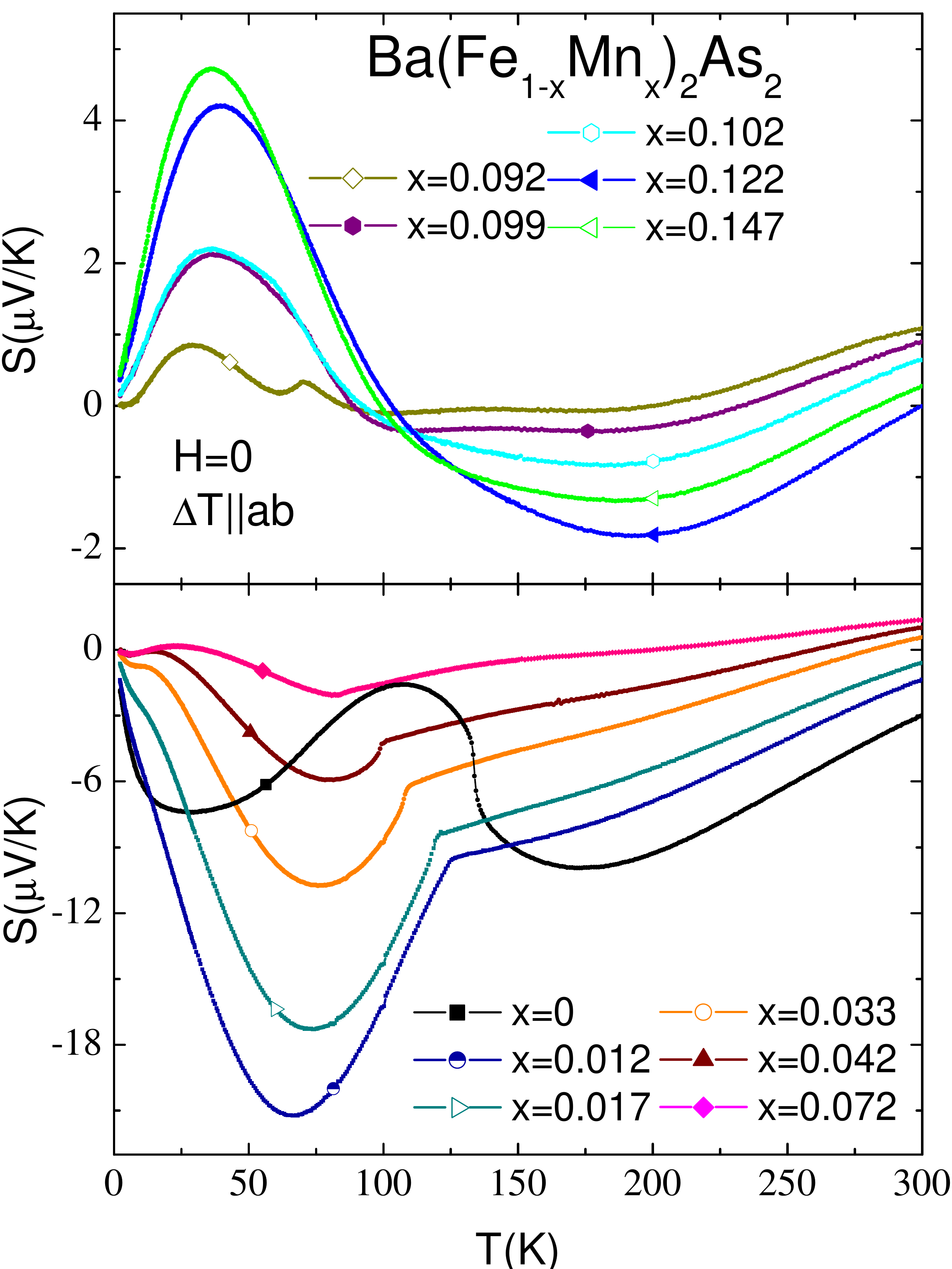}
	\end{center}
	\caption{In-plane $S(T)$ of Ba(Fe$_{1-x}$Mn$_x$)$_2$As$_2$) (0.026$\le$x$\le$0.147) single crystals.(Color online)}
	\label{fig:TEPoverview}
\end{figure}

Thermoelectric power has proven to be a very sensitive probe of substitution induced changes in BaFe$_2$As$_2$ based materials.\cite{munTEP,halynaTEP} Fig.~\ref{fig:TEPoverview} shows the thermoelectric power ($S(T)$) as a function of temperature from 2~K to 300~K. The thermoelectric power data clearly show an anomaly associated with the structural/magnetic transition that can be tracked as a function of $x$.\cite{munTEP,halynaTEP} The low-concentration data show a kink associated with the structural/magnetic transition. As the concentration of Mn increases, the transition is suppressed, the kink flattens out and the structural/magnetic transition is more clearly seen in the $dS/dT$ plots (Fig.~\ref{fig:TEPcrit}). For $x<0.1$, the structural/magnetic transition temperatures extracted from $S(T)$ show good agreement with those obtained from magnetization and resistivity measurements (Fig.~\ref{fig:XvTPD} below).



Another noteworthy trend revealed in Fig.~\ref{fig:TEPoverview} is that as $x$ increases there is a shift from purely negative values of $S(T<300K)$ for $x<0.033$ to increasingly positive values of $S(T\sim50K)$ for $x\gtrsim0.10$. These results imply that there may be a change in the sign of the charge of the dominant carriers as $x$ increases, at least at low temperature.

\section{discussion}

\begin{figure}[h]
	\begin{center}
		\includegraphics[width=.9\linewidth]{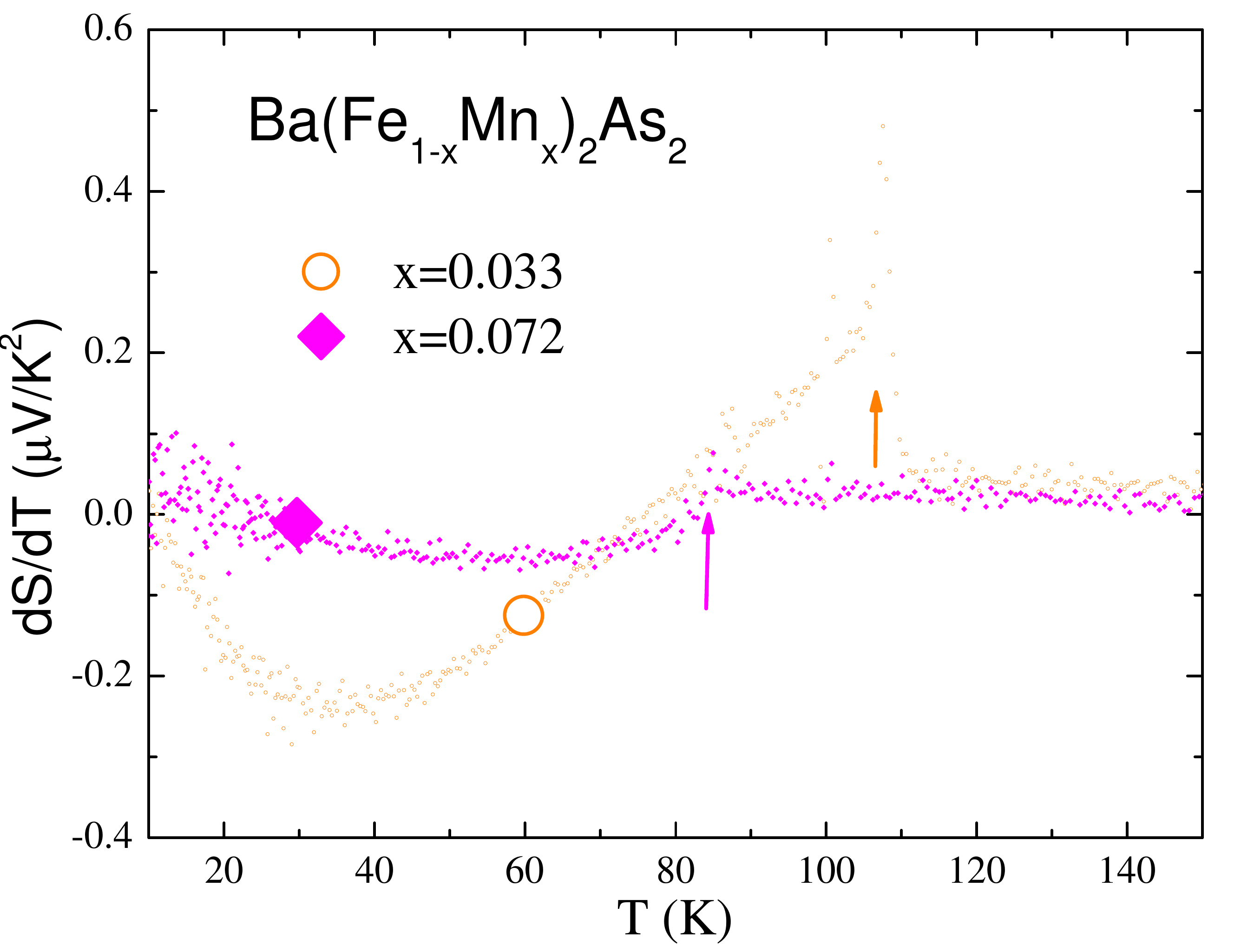}
	\end{center}
	\caption{Derivatives of TEP for representative concentrations. $T_s/T_m$ marked with arrows. (Color online)}
	\label{fig:TEPcrit}
\end{figure}

\begin{figure}[h]
	\begin{center}
		\includegraphics[width=.9\linewidth]{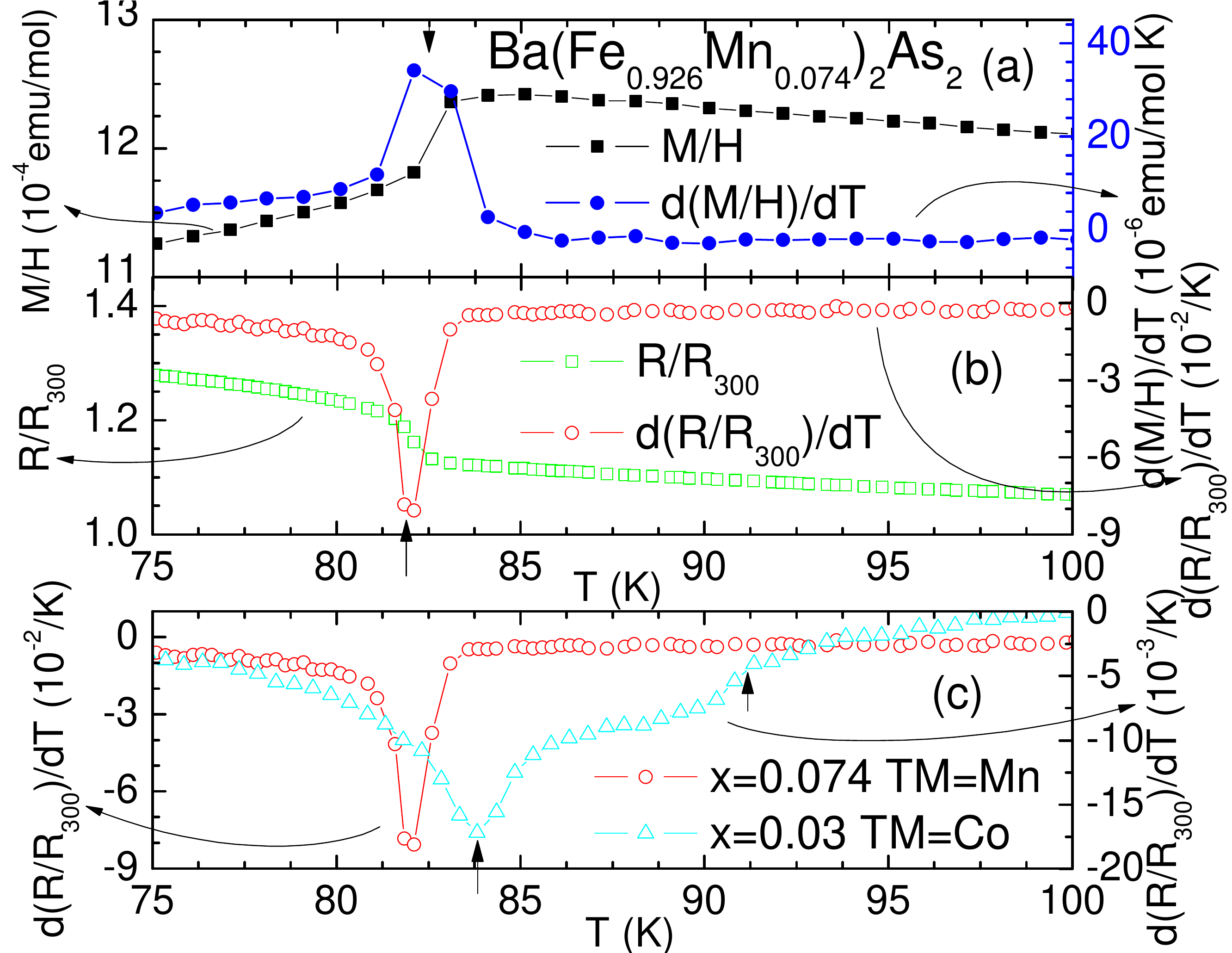}
	\end{center}
	\caption{Magnetization (a) and resistivity (b), along with derivatives, for Ba(Fe$_{1-x}$Mn$_x$)$_2$As$_2$ ($x=0.074$).  Vertical arrows show criterion for transition temperature. (c) shows resistivity derivative data for Co substitution (x=0.03) with a similar transition temperature, with the vertical arrows indicating the higher temperature structural and lower temperature magnetic transitions.  (Color online)}
	\label{fig:KQ905transplot}
\end{figure}

Figures~\ref{fig:KQ905transplot}(a) and (b) show magnetization data with its derivative and normalized resistance with its derivative, respectively, for one substitution level ($x=0.074$). The maximum (minimum) of a gaussian fit of the feature in the derivative, shown with an arrow, gives the value of the transition temperature used throughout this work. The uncertainty is taken as the full-width, half-max of this gaussian fit. Fig.~\ref{fig:TEPcrit} shows $dS/dT$ data for a nearby substitution level ($x=0.072$) as well as the criterion used to determine the transition temperature from thermopower data.

\begin{figure}[h]
	\begin{center}
		\includegraphics[width=.9\linewidth]{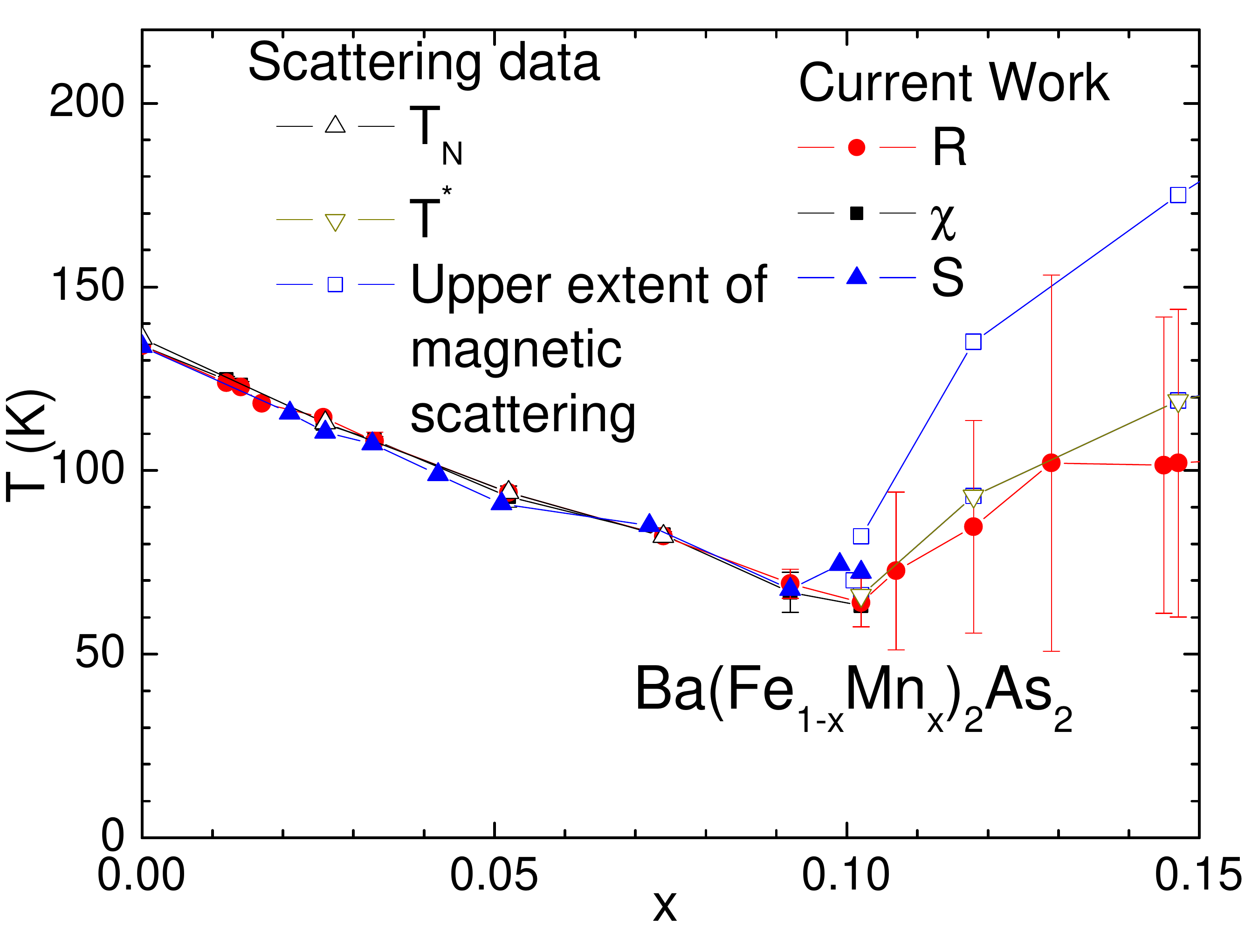}
	\end{center}
	\caption{$T-x$ phase diagram for Ba(Fe$_{1-x}$Mn$_x$)$_2$As$_2$ single crystals for $0<x<0.2$.  Also shown is data from neutron scattering for comparison.\cite{kimMn} For $x\gtrsim0.1$ the transition temperature inferred from the broad resistive feature roughly agrees with the temperature ($T^*$) below which neutron scattering detects long range magnetic order.\cite{kimMn} (Color online)}
	\label{fig:XvTPD}
\end{figure}

As seen in Fig.~\ref{fig:XvTPD}, a T-x phase diagram for the Ba(Fe$_{1-x}$Mn$_{x}$)$_2$As$_2$ system has been constructed by applying these criteria to the data from each member of the substituted series.
Note that whereas a resistive feature is seen for all $x<0.15$, the magnetic and thermoelectric transition signatures are absent for $x>0.1$.
Substitution of Mn for Fe suppresses the high temperature tetragonal to orthorhombic and antiferromagnetic phase transition, bringing it to a clear minimal value of $\sim65$~K for $x\sim0.1$, at the approximate rate of 7~K$/\%\text{Mn}$. Of primary interest are the absence of superconductivity and the lack of split between structural and magnetic phase transition temperatures. This latter point is most clearly illustrated in Fig.~\ref{fig:KQ905transplot}(c), where we show resistive derivatives of both Mn and Co substituted samples with similar transition temperatures. The data from the Co substituted sample show two distinct features: a sharp minimum which has been associated with the magnetic transition and a somewhat higher temperature "shoulder" associated with the structural transition. By contrast, the data from the Mn substituted sample shows only the sharp minimum. This is similar to the parent compound, where the transitions are seen at essentially the same temperature (134 K). This clear lack of splitting for $x<0.1$ is conclusively demonstrated by neutron and x-ray scattering measurements.\cite{kimMn}

There is excellent agreement between different measurement techniques for $x<0.1$: $T_{s/m}$ as measured by resistance, magnetization, thermopower and scattering techniques (Ref.~\onlinecite{kimMn}) are very close (less than 1~K difference). For $x>0.1$, the features are broadened.
The temperature associated with the broad minimum in the temperature dependent derivative of the resistance agrees well with the temperature associated with the onset of long range magnetic order ($T^*$ in Fig.~\ref{fig:XvTPD}) and for several series members with $0.10<x<0.15$ a small shoulder-like feature can be detected near the temperature that magnetic scattering is first detected (Fig.~\ref{fig:XvTPD}).\cite{kimMn}

\begin{figure}[h]
	\begin{center}
		\includegraphics[width=.9\linewidth]{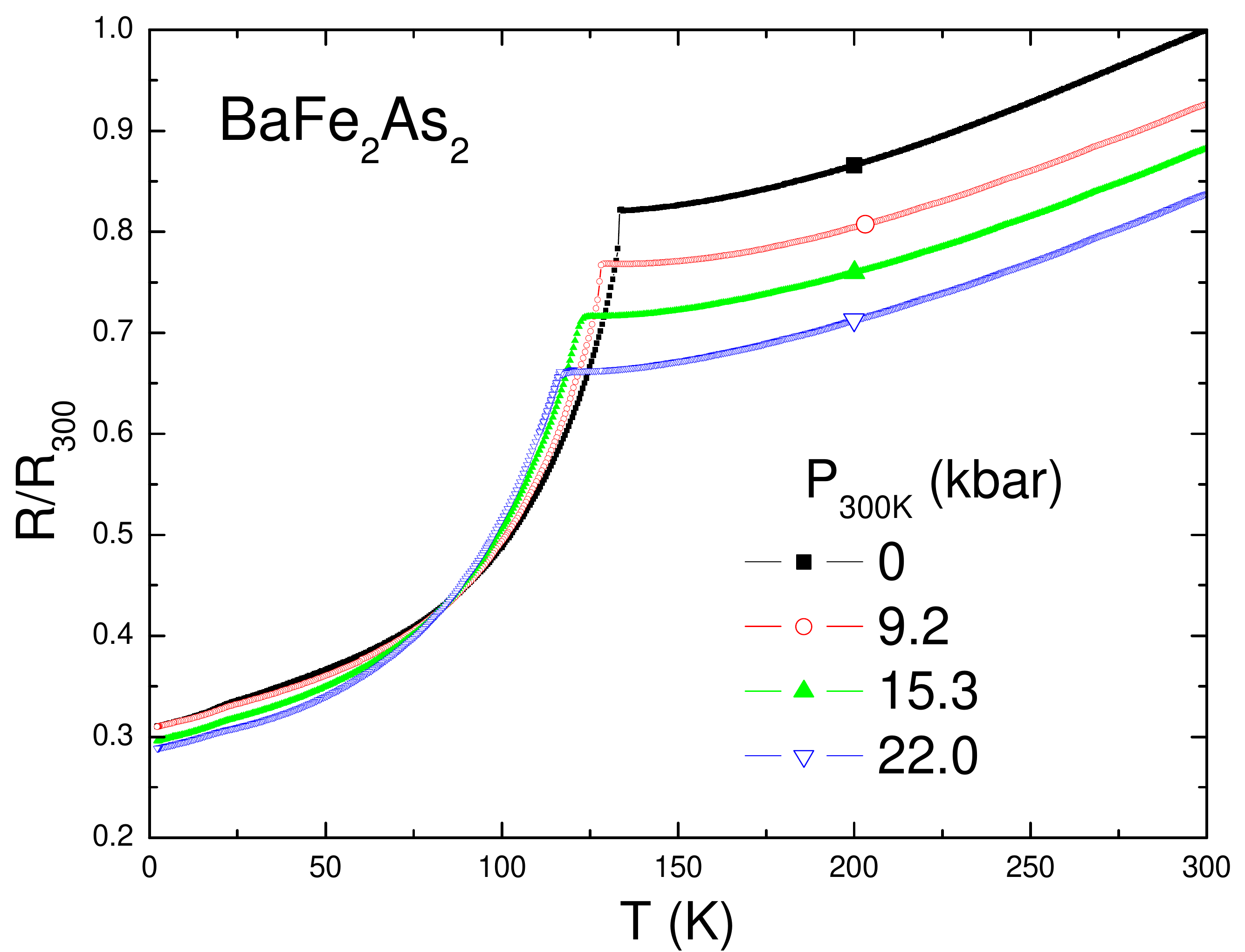}
		\includegraphics[width=.9\linewidth]{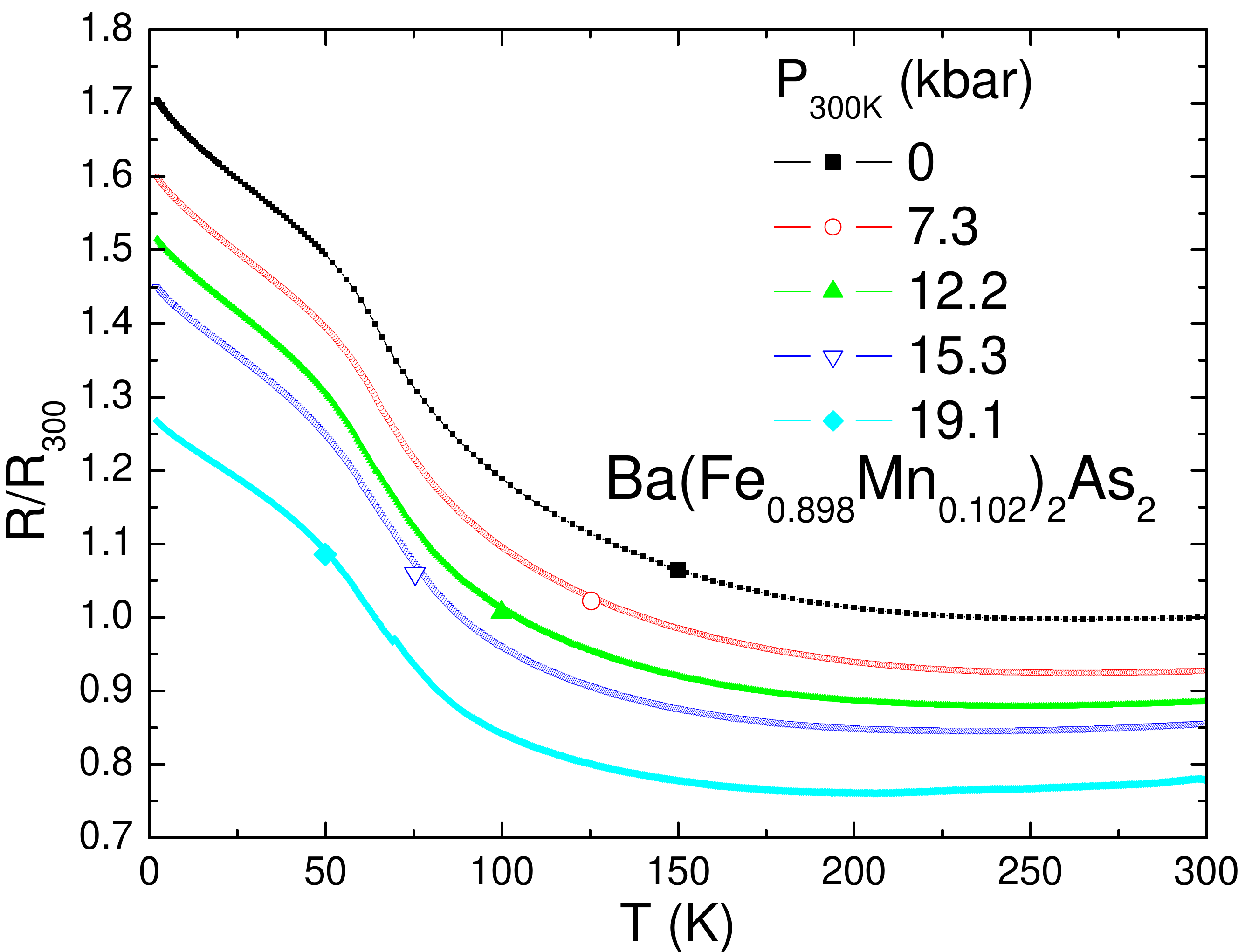}
	\end{center}
	\caption{Top: electrical resistivity, normalized to the ambient pressure, room temperature value, vs temperature for the parent compound BaFe$_2$As$_2$ in pressures up to 22~kbar. The inset shows the pressure dependence of the temperature of the magnetic/structural transition. Bottom: electrical resistivity, normalized to the ambient pressure, room temperature value, vs temperature for the compound with $x=0.102$ Mn substitution, showing a drop in resistivity under pressure over the entire temperature range, without any significant effect on the structural transition. (Color online)}
	\label{fig:pressure}
\end{figure}

\begin{figure}[h]
	\begin{center}
		\includegraphics[width=.9\linewidth]{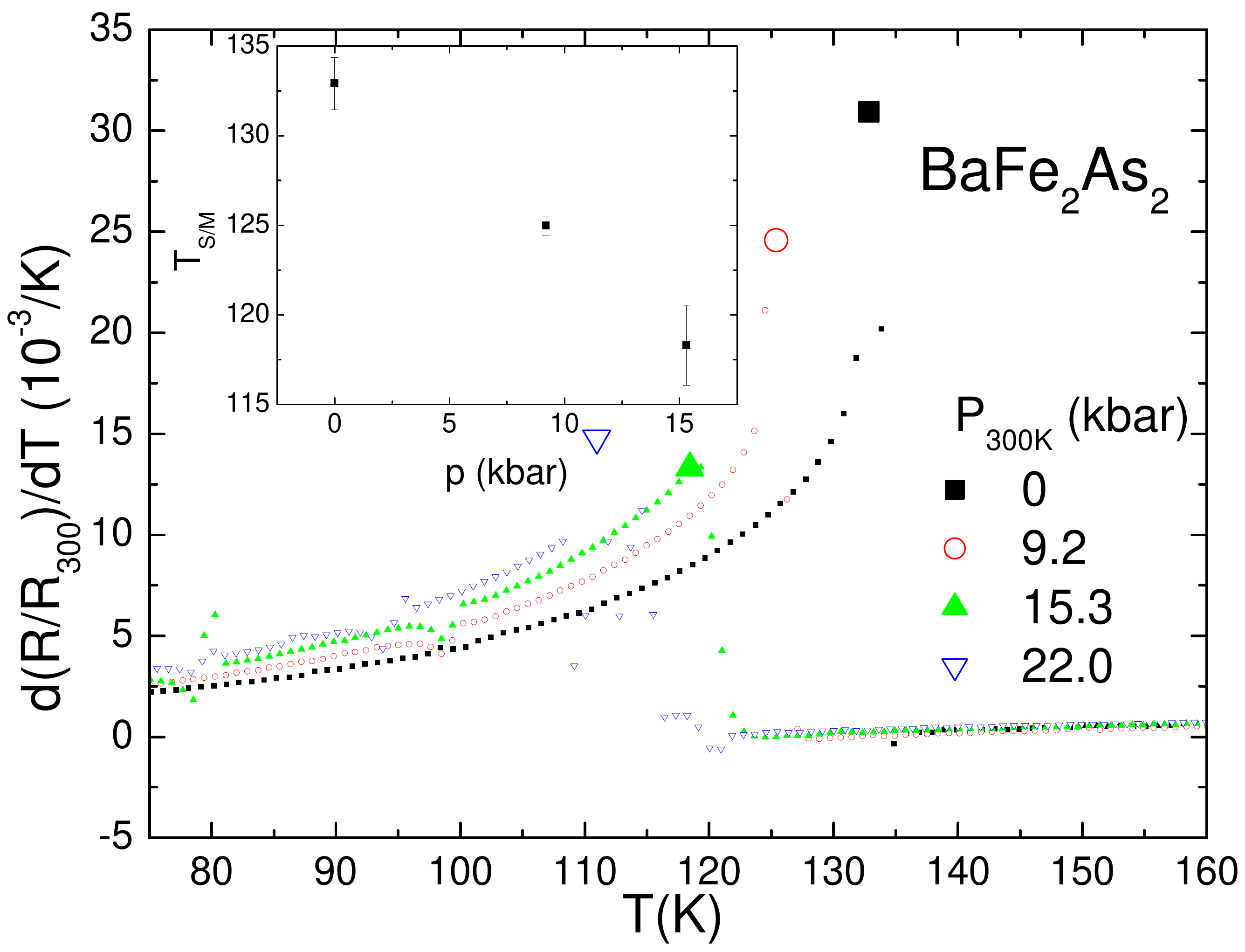}
		\includegraphics[width=.9\linewidth]{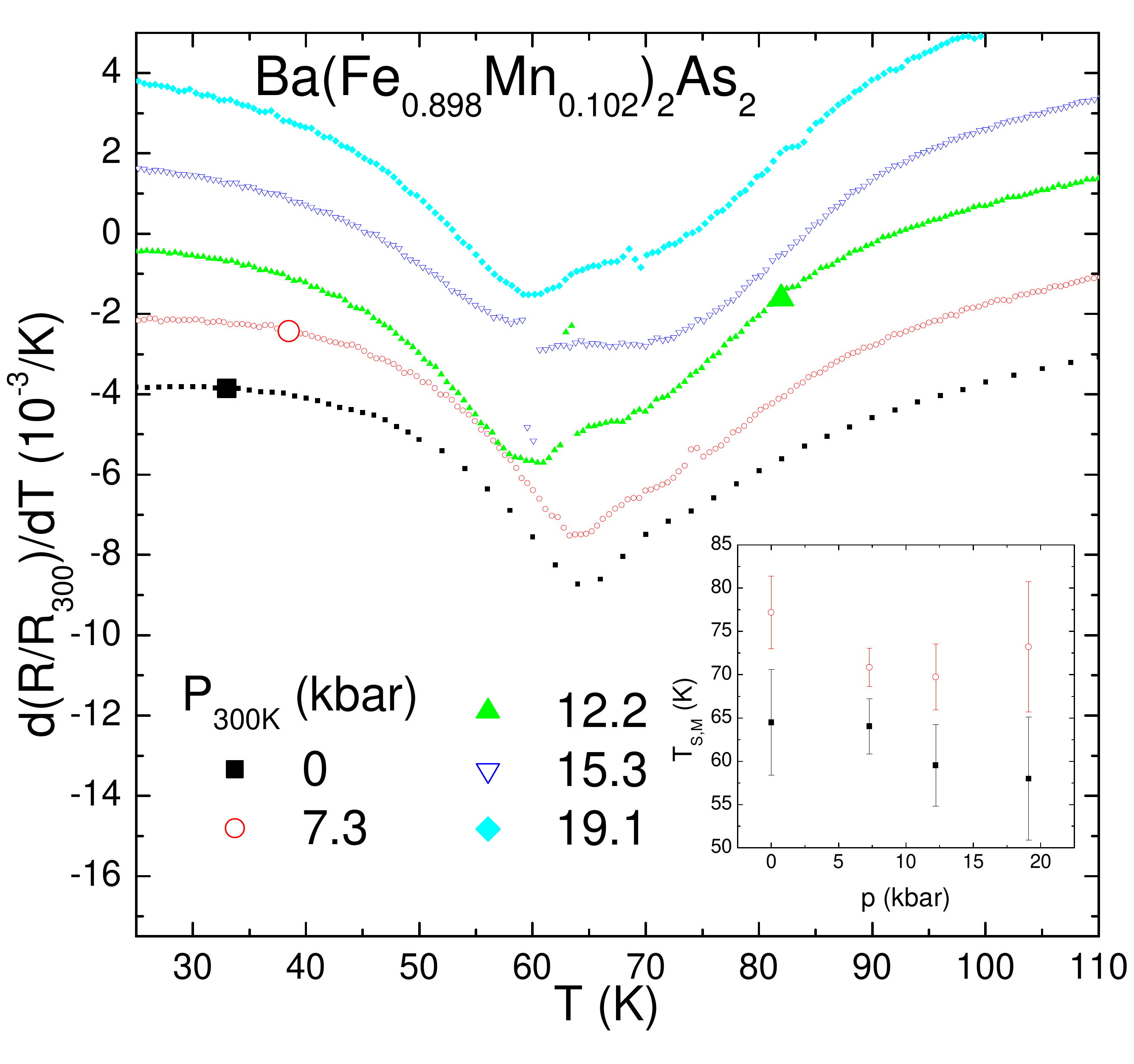}
	\end{center}
	\caption{Top: derivative of normalized electrical resistivity vs temperature for the parent compound BaFe$_2$As$_2$ in pressures up to 22~kbar. Bottom: derivative of normalized electrical resistivity vs temperature for the compound with $x=0.102$ Mn substitution. Note that an artificial cumulative offset of $2\times 10^{-3}$ has been applied to each series in the bottom plot to aid in visibility. (Color online)}
	\label{fig:pressurederiv}
\end{figure}

$x=0.102$ is a special point in the phase space, it is the concentration with the lowest value for $T_{s/m}$ and
is the concentration at which an apparently different type of behavior is emerging (Ref.~\onlinecite{kimMn}) and, as such, is worth further investigation. The effects of hydrostatic pressures up to $\sim$20~kbar on the behavior of the $\rho$~vs~T for the parent BaFe$_2$As$_2$ and $x=0.102$ Mn substituted compounds are shown in Fig.~\ref{fig:pressure}. For the parent compound, pressure reduces $T_{s/m}$ at the approximate rate of $-0.92$~K/kbar; while the change in $\rho$ below $T_{s/m}$ is small, pressure reduces $\rho$ for $T$~$>$~$T_{s/m}$ significantly, e.g.~$d(\rho/\rho_{300K})/dP \approx -1 \%/kbar$. Higher pressures ($\sim 50$ kbar, not shown) lead to the further suppression of the structural transformation and the emergence of superconductivity.\cite{colombier} Pressure reduces $\rho$ of the $x = 0.102$ Mn substituted compound for $T > T_{s/m}$ at a slighly faster rate, e.g.~$d(\rho/\rho_{300K})/dP\approx -1.1 \%/kbar$; however, unlike the parent compound, the values of $\rho$ for $T < T_{s/m}$ are also reduced by pressure. At first glance, the broad S-shaped anomaly in the $x = 0.102$ compound, which presumably reflects the breadth of the structural/magnetic transformation, appears not to be affected significantly by pressure. A closer look at the $d(R/R_{300})/dT$ data for $x=0.102$ (Fig.~\ref{fig:pressurederiv}) reveals a separation or splitting of the transition.
As mentioned above and as shown in Fig.~\ref{fig:XvTPD}, for $x=0.102$ neutron scattering data, as well as ambient pressure resistivity data, detect two features associated with magnetic ordering. The application of pressure sharpens and separates these two features with the lower transition decreasing by roughly $-0.4$~K/kbar and the upper feature manifesting a weaker, but non-monotonic, pressure dependence.
In summary, although separately both pressure and Mn substitution have similar effects in lowering $T_{s/m}$, their effects in the $x = 0.102$ compound are not cumulative.
In addition, the separation of the transitions
with applied pressure is both noteworthy and intriguing.

\begin{figure}[h]
	\begin{center}
		\includegraphics[width=.9\linewidth]{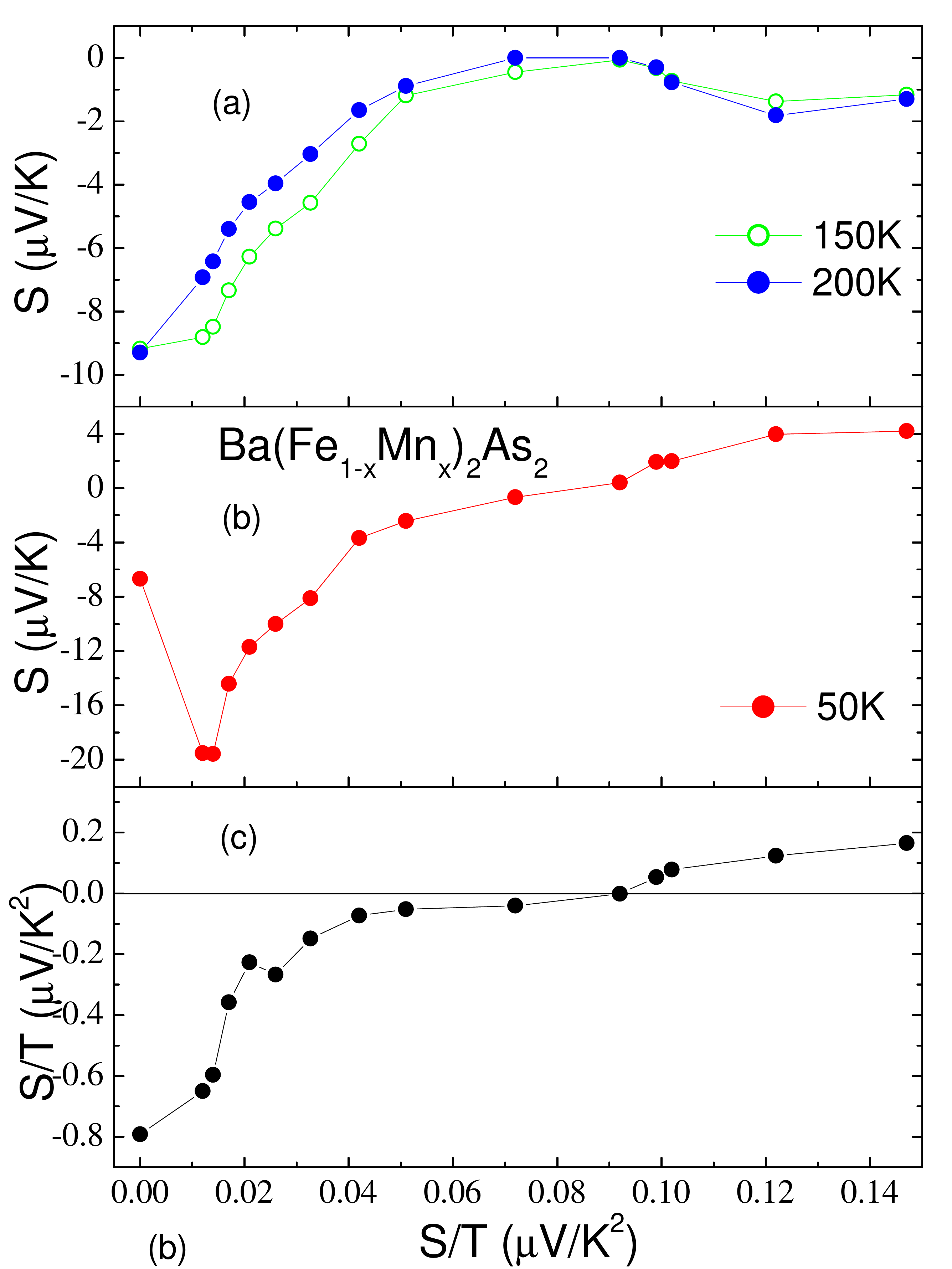}
	\end{center}
	\caption{(a) and (b) TEP as a function of concentration at fixed (150~K, 200~K, and 50~K respectively) temperatures. (c) Low-temperature values of S/T for Ba(Fe$_{1-x}$Mn$_x$)$_2$As$_2$ (0.026$\le$x$\le$0.147).(Color online)}
	\label{fig:lowTTEP}
\end{figure}

In addition to phase diagram determination, $S(T,x)$ data can sometimes identify other changes in electronic structure or character.\cite{varlamov,blanter,behnia}
$S$ as a function of concentrations at fixed temperatures is plotted in Fig.~\ref{fig:lowTTEP}, upper panels, and low temperature, linear in $T$, coefficient of thermopower, $S/T$ as a function of concentration (Ref.~\onlinecite{behnia}) in Fig.~\ref{fig:lowTTEP}, lower panel. The low temperature $S/T$ parameter was determined from a linear fit of the $S(T)$ data below $\sim$4~K.
The only dramatic feature in the data is the sharp change in $S(50\text{~}K)$ for finite values of $x$, but this has been associated with changes in the scattering upon substitution.\cite{thaler,tillman,CanfieldOverview}
Beyond this, a fairly subtle feature can be seen in $S(x)$ between x$\sim$0.092 and x$\sim$0.102 for all plotted isotherms. Whereas for $T=50$ K this could be due to passing through a structural/magnetic transition,
the anomaly seen in $S(x)$ at $T=150$~K and $T=200$~K (temperatures above $T_s/T_m$) at the same concentration is of a different character. In addition, the $\frac{S(x)}{T}$ line crosses zero between x$\sim$0.092 and x$\sim$0.102 (Fig.~\ref{fig:TEPcrit}(c)).
These analyses suggest that a Lifshitz transition, or other change in the electronic structure, may occur in this substitution range.

\begin{figure}[h]
	\begin{center}
		\includegraphics[width=.9\linewidth]{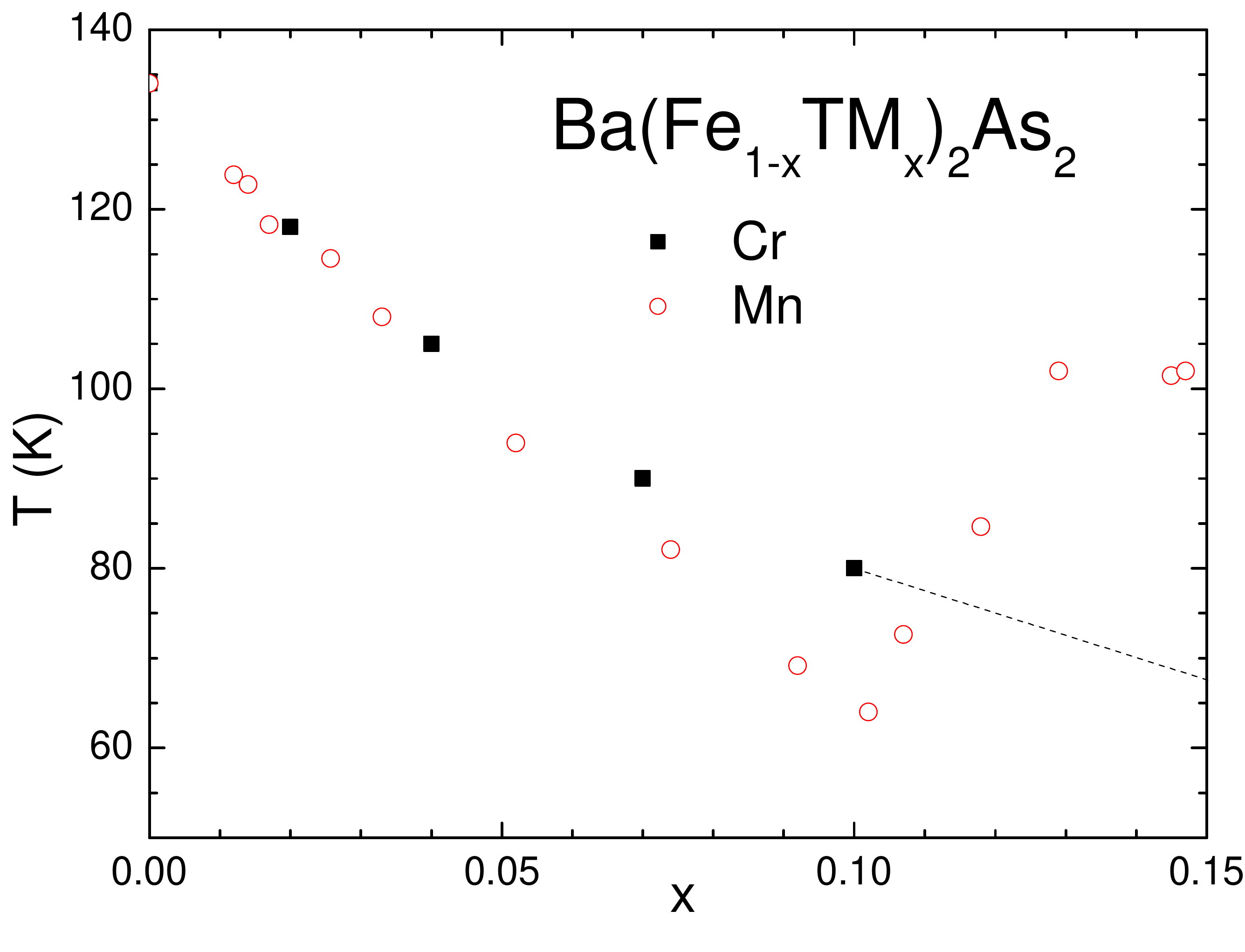}
	\end{center}
	\caption{Comparison of the $T-x$ phase diagrams of Ba(Fe$_{1-x}$Mn$_x$)$_2$As$_2$ and Ba(Fe$_{1-x}$Cr$_x$)$_2$As$_2$ (Ref.~\onlinecite{sefatCr}) for $0$~$<$~$x$~$<$~$0.15$. The dashed line connects to the next higher substitution concentration in the Cr series ($x=0.18$). (Color online)}
	\label{fig:CrMnPDcompare}
\end{figure}

The effects of Mn substitution can be compared with those of
Cr substitution.
Both substitutions increase both the $a$ and $c$ lattice parameters with $x$ (Ref. \onlinecite{sefatCr}), but at different rates.
At $x\sim 0.1$, the change in unit cell volume for both Mn and Cr substitutions is about $0.8\%$, even though the change in $a$ is about $0.05\%$ in Cr and $0.3\%$ in Mn and the change in $c$ is $0.7\%$ in Cr and $0.2\%$ in Mn.
Figure~\ref{fig:CrMnPDcompare} shows a comparison of the T-x phase diagrams of the two compounds. Below $x\sim0.1$, they are quite similar: both display suppression of $T_{s/m}$ to similar degrees, and neither shows any evidence of superconductivity. Above $x\sim0.1$ however, they begin to differ: in contrast to the full suppression of the structural transition in Ba(Fe$_{1-x}$Mn$_x$)$_2$As$_2$ above $x\sim0.1$ (Ref.~\onlinecite{kimMn}), the structural transition is observed in Ba(Fe$_{1-x}$Cr$_x$)$_2$As$_2$ up to at least $x\approx0.18$.\cite{MartyCr} In addition, the magnetic transition temperature monotonically decreases over the entire range in the Cr substituted series, (Refs.~\onlinecite{sefatCr,MartyCr}), in contrast with the upturn seen in Mn substitution.

\begin{figure}[h]
	\begin{center}
		\includegraphics[width=.9\linewidth]{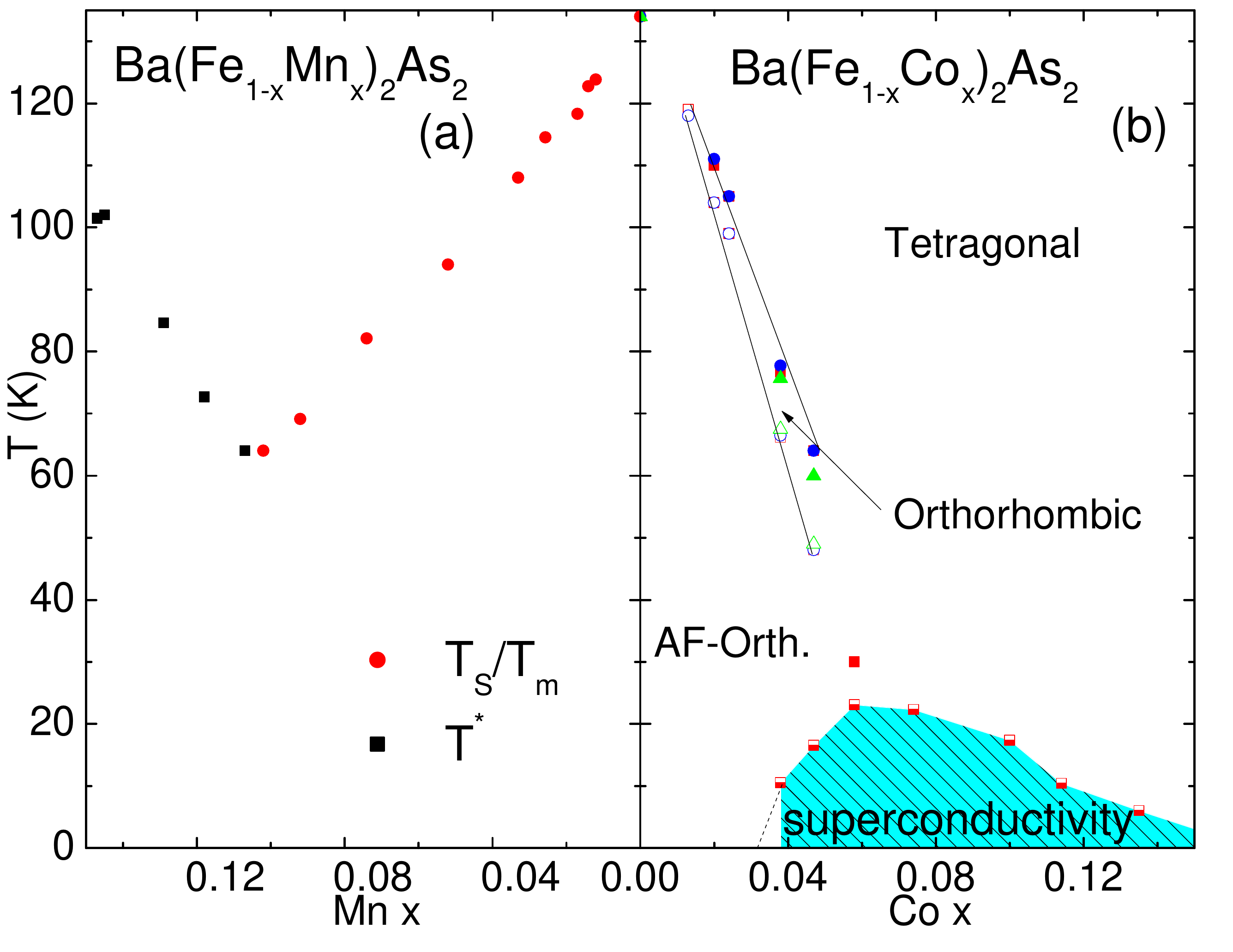}
	\end{center}
	\caption{(a) Ba(Fe$_{1-x}$Mn$_x$)$_2$As$_2$ $T-x$ phase diagram. (b)~Ba(Fe$_{1-x}$Co$_x$)$_2$As$_2$ phase diagram.\cite{tillman} (Color online)}
	\label{fig:PDwithCo}
\end{figure}

Fig.~\ref{fig:PDwithCo} highlights some of the key differences between Mn and Co substitution for $x<0.15$. Of primary interest is the difference in the slope of the suppression of $T_{s/m}$. When comparing Co, Ni and Cu substitution, the suppression is seen to be dependent on atomic substitution concentration regardless of which substitution is used, and is thus not seen to be dependent on the number of extra carriers.\cite{canfield} By contrast, the slope with Mn (or Cr) substitution is about half as steep.
It is unclear if this is due to the difference between electron and hole substitution or other differences such as structural changes. In addition, for Co, Ni, Rh, Pd and Cu, substitution appears to split the structural and magnetic transition temperatures (Ref.~\onlinecite{CanfieldOverview}), $T_s$ and $T_m$, with a maximum split of 16 K at 5.8\% Co, 19 K at 3.2\% Ni and 15 K at 3.5\% Cu.\cite{BigNiDoping}
For Mn substititution though, there is no resoluble splitting in the transport, thermodynamic or microscopic (Ref.\onlinecite{kimMn}) data for $x\lesssim 0.1$.

Taking the minimum value for $T_{s/m}\approx65 $~K at $x\sim 0.1$ in the Ba(Fe$_{1-x}$Mn$_x$)$_2$As$_2$ system and comparing it to the Ba(Fe$_{1-x}$Co$_x$)$_2$As$_2$ series, we get that the equivalent Co substitution concentration for the same magnetic transition temperature is $x\sim 0.04$. This is on the low substituted side of the superconducting dome, but with a well defined $T_c\approx 11K$.\cite{tillman} We clearly do not see any sign of such superconductivity at this Mn composition. This indicates that the well defined scaling of $T_c$ with suppression of $T_s$ or $T_m$ (Refs.~\onlinecite{canfield,thaler,CanfieldOverview}) that is seen for TM~=~Co,~Ni,~Rh, and Pd substitution does not hold for TM~=~Mn.

\section{Summary}
Single crystals of Ba(Fe$_{1-x}$Mn$_x$)$_2$As$_2$ can be flux grown for a continuous range of substitution concentrations from $0<x<0.15$. Samples with homogeneous, single phase concentrations above this can not easily be grown, as
the crystals phase separate into Mn- and Fe-rich mesoscopic regions at
intermediate substitutions. As $x$ increases from zero up to $x\approx0.1$, the structural and magnetic phase transition temperature, $T_s/T_m$, is suppressed but does not clearly split as it does for TM~=~Co,~Ni,~Cu,~Rh, and Pd substitution.  For $x>0.1$, a resistive feature associated with magnetic ordering
is observed with a transition temperature that increases with $x$.  Superconductivity is not observed for any value of $x$.  The suppression of 
$T_s/T_m$ occurs at a slower rate for Mn substitution than for substitution with TM~=~Co, and slightly more quickly than for TM~=~Cr.

\section{Acknowledgements}
Work at the Ames Laboratory was supported by the Department of Energy, Basic Energy Sciences under Contract No. DE-AC02-07CH11358. Pressure measurements were supported by the National Science Foundation under Grant No. DMR-0805335. We would like to thank S. Bud'ko, A. Kreyssig, S. Kim, X. Lin, R. Hu, E. Colombier, W. McCallum, K. Dennis and M. G. Kim for help and useful discussions.

\bibliography{mnpaper}

\end{document}